\documentclass[a4paper,fleqn,usenatbib]{mnras}

\usepackage[T1]{fontenc}
\usepackage{ae,aecompl}

\usepackage{graphicx}	\usepackage{amsmath}	\usepackage{amssymb}

\newcommand{\pcm}{\,pc \,cm$^{-3}$ }	\newcommand{\cm}{\,cm$^{-3}$ }

\title[Mapping interstellar scattering]{Locating the intense 
  interstellar scattering towards the inner Galaxy}

\author[J. Dexter et al.]{
J. Dexter$^{1,2}$\thanks{E-mail: jdexter@mpe.mpg.de}, A. Deller$^{3,4}$,
G.C. Bower$^{5}$, P. Demorest$^{6}$, M. Kramer$^{7,8}$,\newauthor B.W. Stappers$^{8}$, 
A.G. Lyne$^{8}$, M. Kerr$^{9}$, L.G. Spitler$^{7}$, D. Psaltis$^{10,11}$,\newauthor 
M. Johnson$^{12}$, R. Narayan$^{12}$
\\
$^{1}$Max Planck Institute for Extraterrestrial Physics,
Giessenbachstr. 1, 85748 Garching, Germany\\
$^{2}$Kavli Institute for Theoretical Physics, University
of California, Santa Barbara, CA 93106, USA\\
$^{3}$ASTRON, the Netherlands Institute for Radio Astronomy, Postbus
2, 7990 AA, Dwingeloo, The Netherlands\\
$^{4}$Centre for Astrophysics and Supercomputing, Swinburne University
of Technology, PO Box 218, Hawthorn, VIC 3122, Australia\\
$^{5}$Academia Sinica Institute of Astronomy and Astrophysics, 645
N. A'ohoku Place, Hilo, HI 96720, USA\\
$^{6}$National Radio Astronomy Observatory, Socorro, New Mexico 87801,
USA\\
$^{7}$Max Planck Institute for Radio Astronomy, Auf dem H\"{u}gel 69,
53121, Bonn Germany\\
$^{8}$Jodrell Bank Centre for Astrophysics, University of Manchester, Manchester, M13 9PL, UK\\
$^{9}$Space Science Division, Naval Research Laboratory, Washington, DC 20375-5352, USA\\
$^{10}$University of Arizona, 933 N. Cherry Ave, Tucson, AZ 85721,
USA\\
$^{11}$Radcliffe Institute for Advanced Study and Black-Hole Initiative, Harvard University, 10 Garden St., Cambridge, MA 02138\\
$^{12}$Harvard-Smithsonian Center for Astrophysics, 60 Garden Street, Cambridge, MA 02138, USA\\
}

\date{Accepted XXX. Received YYY; in original form ZZZ}

\pubyear{2015}

\begin{document}
\label{firstpage}
\pagerange{\pageref{firstpage}--\pageref{lastpage}}
\maketitle

\begin{abstract}
We use VLBA+VLA observations to measure the sizes of the
scatter-broadened images of 6 of the most heavily scattered known 
pulsars: 3 within the Galactic Centre (GC) and 3 elsewhere in the
inner Galactic plane ($\Delta l < 20^\circ$). By combining the measured sizes with temporal pulse
broadening data from the literature and using the thin-screen
approximation, we locate the scattering medium 
along the line of sight to these 6 pulsars. At least two 
scattering screens are needed to explain the observations of the GC
sample. We show that the screen inferred by previous 
observations of SGR J1745$-$2900 and Sgr~A*, which must be located far
from the GC, falls off in strength on scales
$\lesssim 0.2$ degree. A second
scattering component closer to ($\Delta < 2$ kpc) or even
(tentatively) within ($\Delta < 700$ pc) the GC produces most or all of the
temporal broadening observed in the other GC pulsars. Outside the GC,
the scattering locations for all three pulsars are $\simeq 2$ kpc from Earth,
consistent with the distance of the Carina-Sagittarius or Scutum spiral arm. For
each object the 3D scattering origin coincides with a known HII region
(and in one case also a supernova remnant), suggesting that such
objects preferentially cause the intense interstellar scattering seen towards
the Galactic plane. We show that the HII regions should contribute
$\gtrsim 25\%$ of the total dispersion measure (DM) towards these
pulsars, and calculate reduced DM distances. Those distances for other pulsars
lying behind HII regions may be similarly overestimated. 
\end{abstract}


\section{Introduction}

Interstellar scattering from electron density inhomogeneities leads to
multipath propagation, broadening the radio 
images and pulse profiles of objects in or behind the Galactic
plane. Along certain lines of sight, the scattering is ``intense'' -- 
much larger than predicted by the large-scale components of the Galactic
electron distribution \citep{taylorcordes1993,cordeslazio2002}. It has
long been associated with HII regions and/or supernova remnants near
the line of sight \citep[e.g.,][]{litvak1971,little1973,dennisonetal1984}. 

The Galactic Centre black hole, Sgr~A*, provides a well known example of intense
scattering. Its image size increases as $\lambda^2$ in the radio 
\citep[e.g.,][]{daviesetal1976,backer1978,boweretal2006} as
predicted for a ``thin'' scattering medium
\citep[e.g.,][]{ishimaru1977,blandfordnarayan1985}. The large angular
size of Sgr~A* was previously thought to come from the hot, 
dense gas in the Galactic Centre (GC) region. Producing the large observed image close to
the source would require a special scattering geometry
\citep[][]{goldreichsridhar2006}. It would also prevent the detection
of pulsed radio emission from neutron stars in the GC, potentially explaining the lack of pulsar
detections in the central parsec of the Galaxy \citep{cordeslazio1997,laziocordes1998}.

Radio pulsations discovered from the GC magnetar SGR
J1745$-$2900 \citep{eatoughetal2013}, only 0.1~pc in projection from
Sgr~A*, were broadened by orders of magnitude less than predicted
\citep{spitleretal2014}. In addition, the image size and shape of the
magnetar match that of Sgr~A*, showing that they share the same
scattering medium \citep{boweretal2014}. The scattering medium towards Sgr~A*
therefore does not prevent the detection of ordinary pulsars at frequencies $\gtrsim
3$ GHz. The known young stars in the central parsec imply a
large population of young pulsars. Assuming that the magnetar scattering medium is representative of the
central parsec, the stated sensitivities of deep radio searches to
date and the lack of detections suggest a ``missing pulsar problem''
in the central parsec (\citealt{johnston1994},
\citealt{macquartetal2010}, \citealt{dexteroleary2014}; but see also
\citealt{chennamangalamlorimer2014,rajwadeetal2016,psaltisetal2016}). 

Combining angular and temporal broadening measures for the same source
gives an estimate for the line of sight distance to the scattering
medium \citep{gwinnetal1993,brittonetal1998}. Using this technique, \citet{boweretal2014} showed that the 
scattering medium towards the magnetar and Sgr~A* is not local to the GC, but rather at a distance
$\simeq 2-3$ kpc from Earth, in the nearby Carina-Sagittarius or
Scutum spiral arm. The chance alignment of an ionized gas cloud with Sgr~A* is
highly unlikely unless such clouds cover a significant fraction of the
Galactic plane.

Maser sources are heavily scatter-broadened out to scales of 
$\simeq 0.5$ degree from Sgr~A* \citep{vanlangeveldeetal1992}. Observations of Sgr~A* and the magnetar show that
the scattering medium extends over scales of arcseconds, but it is
not clear how much if any of the rest of the observed scattering in
the GC (here defined as the central 0.5 deg) has the
same physical origin. The nearest known pulsars to the magnetar are at
separations $\simeq 0.2$ degree
\citep{johnstonetal2006,denevaetal2009detect}. Along with the magnetar, they
are the most temporally broadened known pulsars \citep[e.g.,][]{manchesteretal2005}.

To study the GC screen and the physical origin of intense
scattering, we imaged a sample of strongly scattered pulsars (3 GC, 3 non-GC) with VLBA and
VLBA+VLA observations (\S\ref{sec:observations}). We measure angular
broadening from scattering (\S\ref{sec:size-posit-meas}) in
all sources that we detected (one source was not detected), and locate the scattering
along the line of sight by combining the image sizes with previous
temporal broadening measurements (\S\ref{sec:mapp-strong-interst}). We find evidence for multiple
physical locations for the origin of the GC scattering on $\simeq 0.2$ degree scales,
and tentative evidence for intense scattering local to the GC. We
further show that all three non-GC sources have scattering origins at 
distances $\simeq 2-3$ kpc from Earth like the GC magnetar and Sgr
A*. In all three cases the 3D scattering location coincides with a
known HII region, adding to the evidence that such regions may be the dominant cause of
intense interstellar scattering in the Galactic plane. 

\begin{figure}
\begin{center}
\includegraphics[scale=0.7]{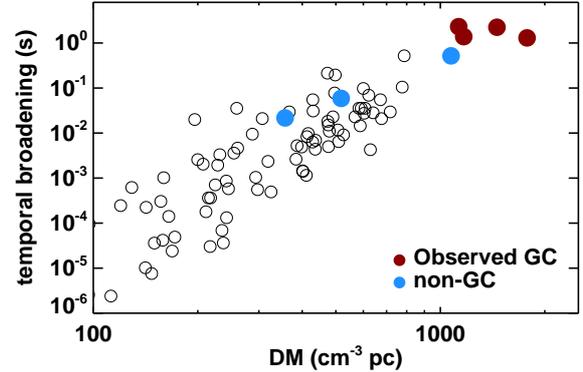}
\caption{\label{taudm}Measured temporal broadening of the pulse
  profile vs. dispersion measure for known pulsars with high DM and measured
  $\tau$ in the ATNF database \citep{manchesteretal2005}. We selected
  accessible objects with the highest possible DM and $\tau$. The
  largest $\tau$ pulsars all reside within the GC (red points),
  including the GC magnetar SGR J1745$-$2900, imaged previously by \citet{boweretal2014}.}
\end{center}
\end{figure}

\begin{table*}
\caption{Some properties of the observed targets and VLBA+VLA observations\label{sources}}
\begin{small}
\begin{center}
\begin{tabular}{lcccccccc}
        \hline
%Name & RA & Dec & l (deg) & b (deg) & DM (pc $\rm cm^{-3}$) & $S_{1400}$ (mJy) & 1 GHz $\tau$ (s)\\
%Name & Date (UT) & Type & $\nu$ (GHz) & Int. time (hr)\\
%        \hline
         Source &     J1745$-$2912 &     J1745$-$2912 &     J1746$-$2849 &     J1746$-$2856 &       B1750$-$24 &       B1758$-$23 &      B1809$-$176 &       B1822$-$14\\
\hline
      l (deg) &  $-0.20$ &  $-0.20$ &    $0.13$ &    $0.12$ &    $4.27$ &    $6.84$ &   $12.91$ &   $16.81$\\
      b (deg) &   $-0.18$ &   $-0.18$ &   $-0.04$ &   $-0.21$ &    $0.51$ &   $-0.07$ &    $0.39$ &   $-1.00$\\
 DM (pc cm$^{-3}$) &      $1130$ &      $1130$ &      $1456$ &      $1168$ &       $672$ &      $1073$ &       $518$ &       $357$\\
        P (s) &    $0.19$ &    $0.19$ &    $1.48$ &    $0.95$ &    $0.53$ &    $0.42$ &    $0.54$ &    $0.28$\\
    S14 (mJy) &     $0.5$ &    $0.5$ &     $0.4$ &     $6.5$ &     $2.3$ &     $2.2$ &     $3.3$ &     $2.6$\\
    Spectral index &   $-1.7$ &   $-1.7$ &   $-1.1$ &   $-2.7$ &   $-1.0$ &   $-1.0$ &   $-1.7$ &   $-1.1$\\
    Obs. Date (UT) &     2015-11-30 &     2016-01-30 &     2015-11-30 &     2016-01-30 &     2015-08-25 &     2015-08-25 &     2015-12-26 &     2015-12-26\\
    Obs. Type &       VLBA+VLA &       VLBA+VLA &       VLBA+VLA &       VLBA+VLA &           VLBA &           VLBA &           VLBA &           VLBA\\
   Calibrator &     J1752-3001 &     J1752-3001 &     J1752-3001 &     J1752-3001 &     J1755-2232 &     J1755-2232 &     J1808-1822 &     J1825-1718\\
    $\nu$ (GHz) &     8.7 &     5.9 &     8.7 &     5.9 &     7.5 &     7.5 &     4.5 &     7.5\\
    Int. time (hr) &    1.84 &    1.58 &    1.52 &    1.51 &    1.75 &
                                                                       1.82
                                                                                &    1.74 &    1.68\\
         Image rms ($\mu$Jy) &     $36$ &    $170$ &     $58$ &     $36$ &    $140$ &    $230$ &    $70$ &     $110$\\
   Beam size (mas) &                     5.0$\times$1.8 &                    14.9$\times$6.3 &                    10.3$\times$4.1 &                    14.6$\times$6.1 &                     7.3$\times$2.3 &                     8.4$\times$4.9 &                    14.1$\times$5.2 &                     8.4$\times$1.8\\
Beam PA (deg) &     $4.0$ &     $2.4$ &     $8.5$ &     $1.7$ &    $-4.2$ &     $7.2$ &   $-10.2$ &   $-13.2$\\
	\hline
\end{tabular}
\end{center}
\end{small}
\end{table*}

\section{Observations and data reduction}
\label{sec:observations}

\subsection{Sample selection}\label{sec:sample-selection}

Our sample was chosen to focus on highly scattered pulsars (dispersion
measure $\rm DM > 200$ \pcm, temporal broadening $\tau > 10$ ms at 1 GHz, which were sufficiently bright for
imaging with the VLBA only (non-GC sources) or the VLBA+VLA (GC sources). 

The sources observed are shown in Figure~\ref{taudm} 
in the $\tau$-DM plane and further properties are listed in Table
\ref{sources}. The four pulsars with the highest measured $\tau$ are
all $\lesssim 0.2$~deg from Sgr A* in the GC \citep{johnstonetal2006,denevaetal2009detect,eatoughetal2013}. The angular
size of the GC magnetar, SGR J1745$-$2900, was measured by
\citet{boweretal2014}. We observed the 3 other sufficiently bright
known GC pulsars. The flat spectrum, young pulsar J1746-2850 was not
observed, since it has not been detected in recent observations
\citep{ngetal2015,schnitzeleretal2016} and may be a magnetar-like
object whose radio emission has since shut off \citep{dexteretal2017}.

Outside of the GC, the four pulsars observed were chosen as those
sufficiently bright to be observable with the VLBA alone and with
suitable nearby ($\le 3$ degrees) calibrators. Most calibrators near
the pulsars are heavily scatter-broadened, and it is often necessary
to go to $\simeq 3$ degrees away to find one that is detectable on
long baselines. The three sources detected (see below) span a range of
an order of magnitude in $\tau$ and a factor of a few in DM.

\begin{figure*}
\begin{center}
\begin{tabular}{lll}
\includegraphics[scale=0.48]{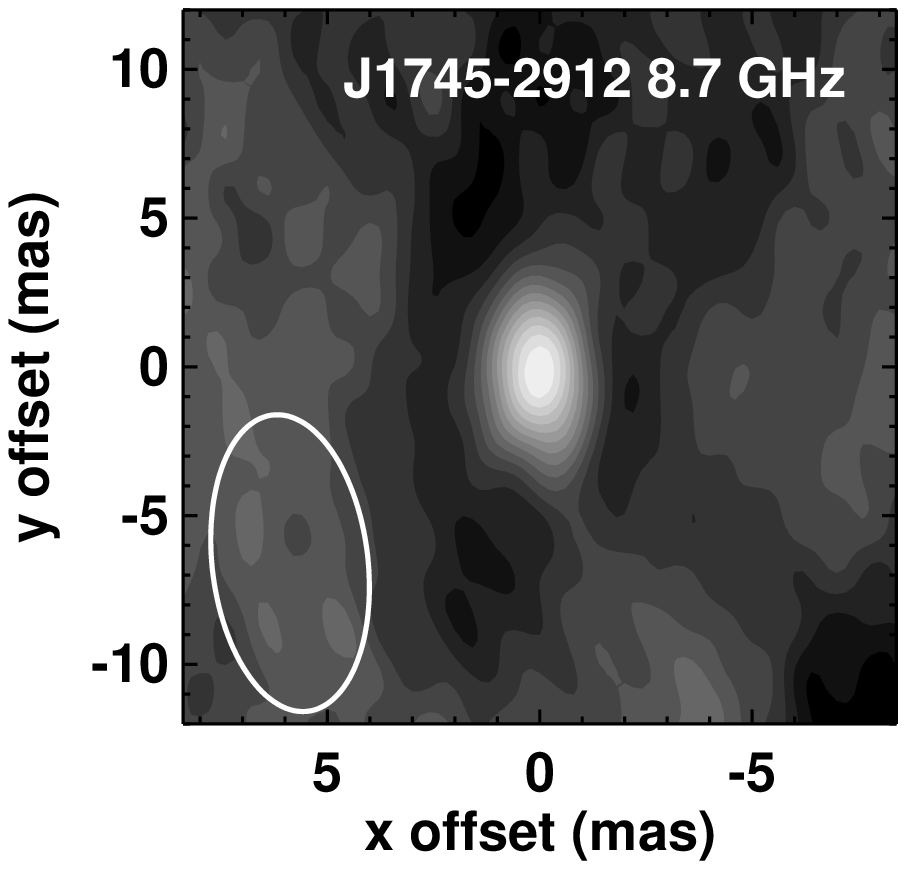}&
\includegraphics[scale=0.48]{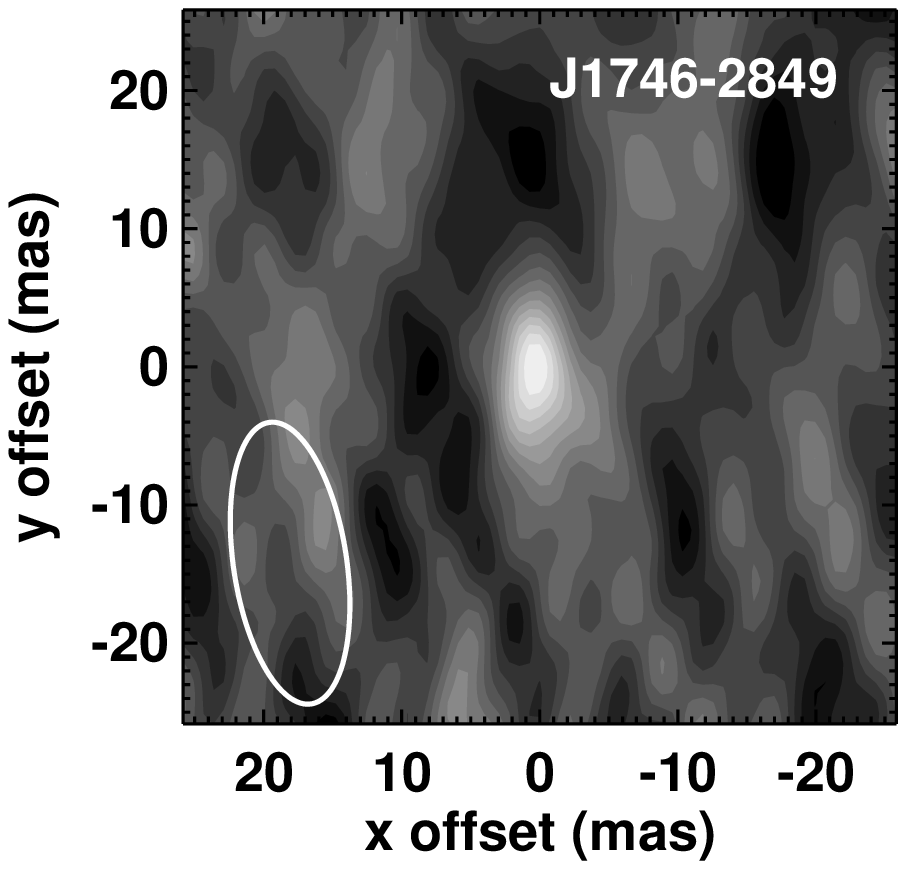}&
\includegraphics[scale=0.48]{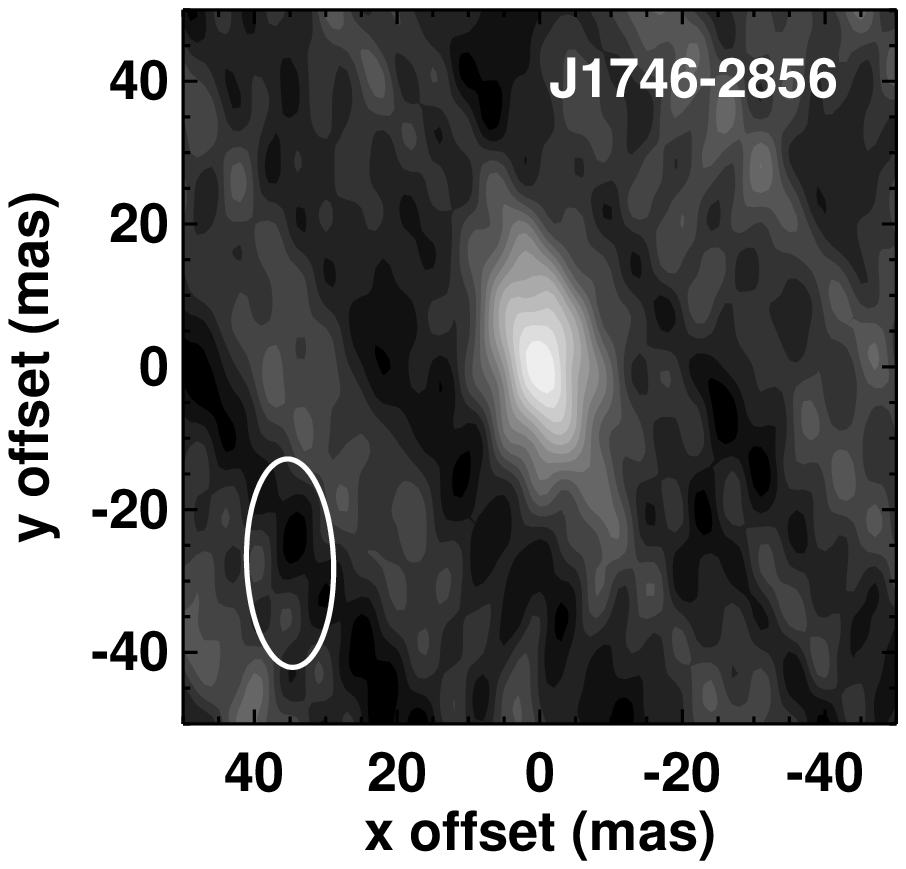}\\
\includegraphics[scale=0.48]{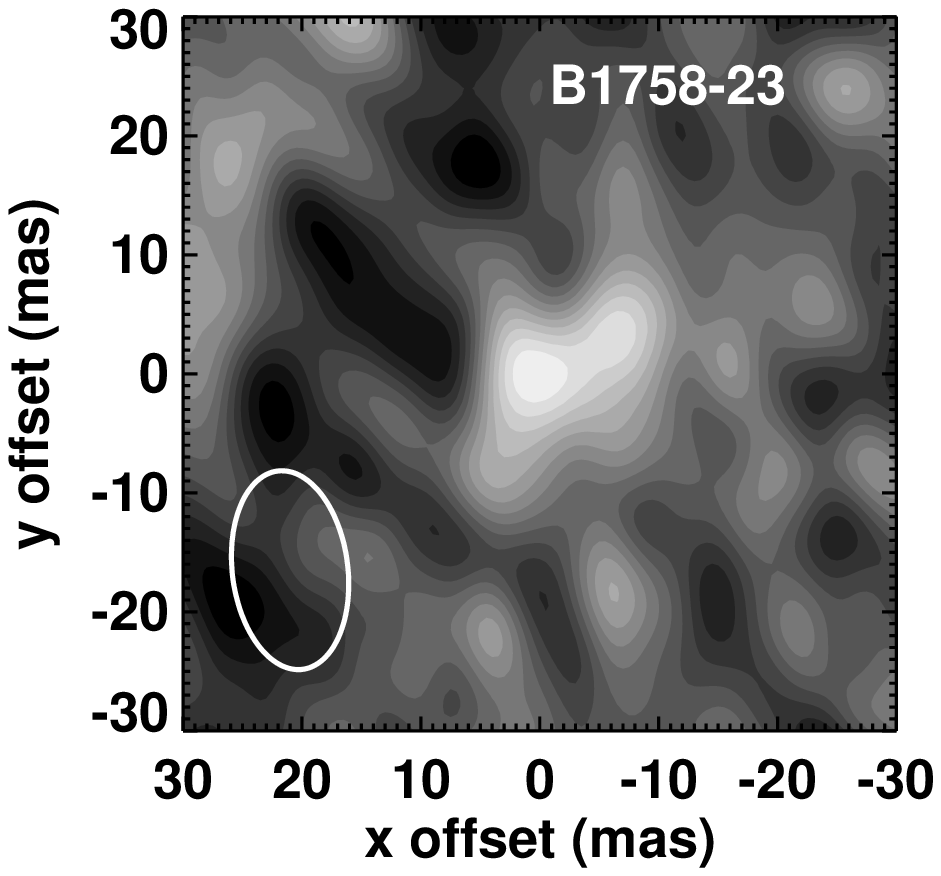}&
\includegraphics[scale=0.48]{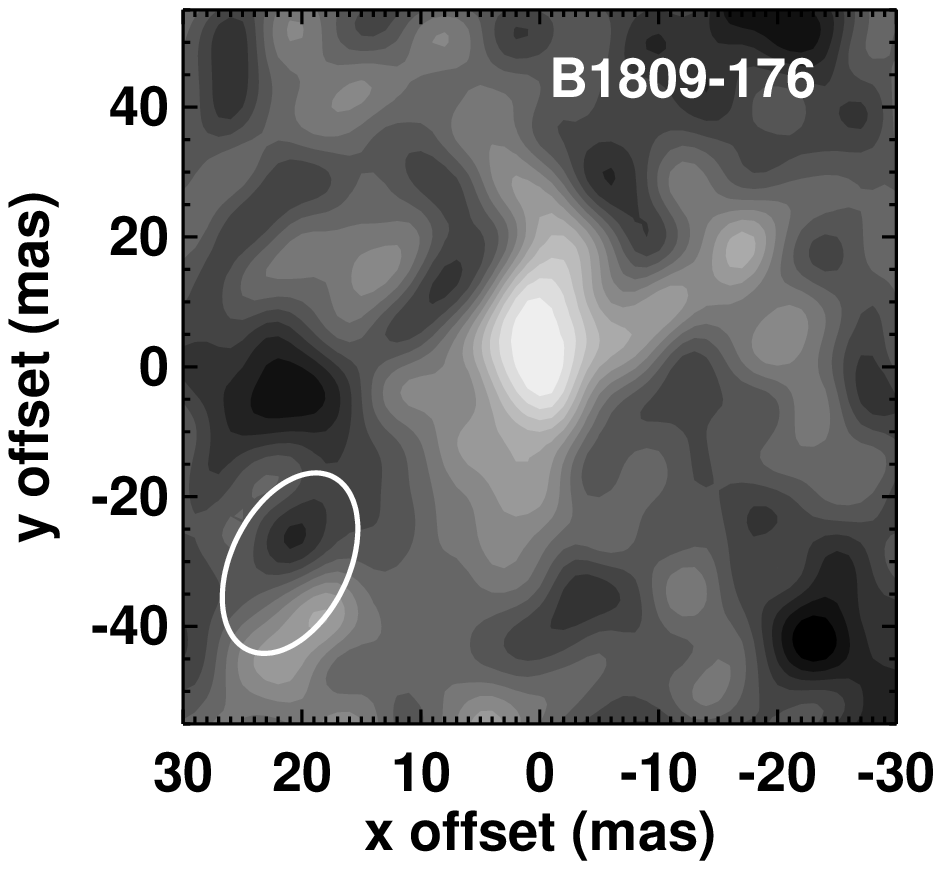}&
\includegraphics[scale=0.48]{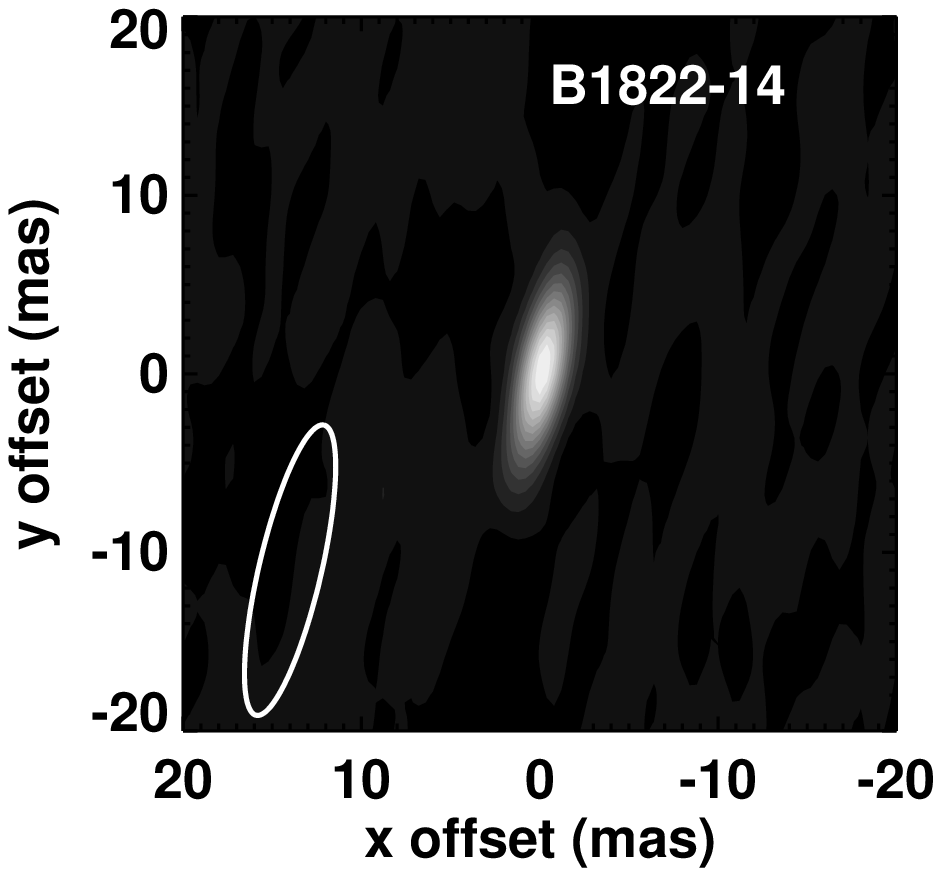}
\end{tabular}
\caption{\label{images}Scatter-broadened images of the 6 detected
  pulsars from our VLBA+VLA sample. The color scale is linear with a
  dynamic range $\simeq 10$. The restoring beam used by CLEAN is shown
as the white ellipse, and is influenced both by the array and source
properties, since in most cases extended source structure leads to
non-detections on long baselines.}
\end{center}
\end{figure*}

\subsection{Observations and correlation}\label{sec:observ-corr}

The parameters of the observations are listed in Table
\ref{sources}. Observing frequencies were chosen by matching the expected scattering
size given the measured temporal broadening and a single scattering
screen at 3 kpc from the Sun to the angular resolution of the inner
six stations of the VLBA. Generally the two sizes are comparable for
$\nu \simeq 4.5-8.7$ GHz. At higher frequency the pulsars are fainter,
while at lower frequency the calibrators are often significantly
scatter-broadened. Pulsar angular sizes significantly smaller
  or larger than expected would appear as unresolved or would not be
  detected. Since we detected 7/8 sources, the choice of observing
  frequencies does not bias our results. The GC pulsars are faint and had large predicted
sizes (implying high frequencies), so that the increased sensitivity
of the VLBA+VLA was needed for detection. In all cases, a data rate of 2 Gbps, corresponding to
256~MHz of bandwidth with dual polarization, was used.  For
observations where the VLA participated, a tied array beam with
filterbank data was produced for all scans on the target pulsars.  

The data were correlated with an integration time of 2 seconds and a
frequency resolution of 0.5 MHz. For the pulsar sources, gating was
employed using ephemerides from timing observations at Jodrell Bank and
Parkes to increase the signal--to--noise ratio.  These ephemerides were
refined using VLA data from the observations themselves where
available, as described below. Amplitude scaling was applied to the
gated data to yield period-averaged equivalent flux densities for the
pulsars, which facilitates comparisons with timing data where the
pulsar flux density is usually quoted in this way.

\subsection{Data calibration and reduction}\label{sec:data-calibr-reduct}

Data reduction was performed with AIPS \citep{greisen2003}, using the
ParselTongue python interface \citep{kettenisetal2006}.  Standard
corrections including {\em a priori} gain calibration based on logged
system temperatures, delay and bandpass calibration on a bright fringe
finder source, and delay, phase, and amplitude calibration on the
phase reference sources were derived.  These cumulative corrections
were applied to the gated data on the target pulsars, before these
data were split and averaged in frequency to a resolution of 32 MHz.
These averaged target datasets were written out in UVFITS format, for
imaging and further analysis as described below. 

\subsection{VLA tied-array data processing}\label{sec:vla-phased-array}

At the VLA it is possible to route the summed-array voltage data stream
to a local compute cluster for real-time detection, integration, and
recording at high time resolution.  For our observing sessions in which
the VLA participated, we enabled this mode in parallel with VLBI
recording in order to obtain simultaneous wide-band timing measurements
of the pulsars to use for gating the VLBI correlation.  These data were
recorded using 1024 MHz total bandwidth, 8-bit voltage quantization,
1024 frequency channels, 0.5~ms time resolution, and summed
polarizations.  The frequency ranges observed were $8.3-9.3$~GHz and
$5.5-6.5$~GHz.  

Offline processing including folding and time-of-arrival measurement was
done using the \textsc{DSPSR} \citep{vanstratenbailes2011} and \textsc{PSRCHIVE}
\citep{hotanetal2004} software pacakges; these data were used to
determine a short-term timing ephemeris (absolute pulse phase and spin
period) that was used to gate the VLBI correlation.  We performed an approximate flux
calibration by scaling the data assuming system equivalent flux
densities for the summed array of 10~Jy and 11.5~Jy, at 9~GHz and 6~GHz
respectively.  The resulting period-averged pulsed flux density
measurements for all pulsars are presented in the far right column of Table \ref{sizetable}. We
conservatively assume $\simeq 50\%$ fractional uncertainty on these
measurements.

\subsection{Source detection}

Imaging was performed using \textsc{difmap} \citep{shepherdetal1994}, employing 
natural weighting for maximum sensitivity to resolved sources. Several targets had significant
positional uncertainties ($\lesssim$ 1 arcsec in dec); we made images minimally
covering a region up to $\pm$3$\sigma$ in R.A. and dec.  After identifying
the pulsar position, we shifted the phase center of the 
visibility data before averaging, to eliminate bandwidth smearing, and
then made the small images centered on the pulsars shown in Figure
\ref{images}. The lowest significance detections have $\simeq 6\sigma$ 
(J1746$-$2849 and B1809$-$176), due to a combination of low total flux
densities ($\sim 0.1-1$ mJy) and resolved sources. 

%\documentclass{article}
%\begin{document}
%\footnotesize
\begin{table*}
\caption{\label{sizetable}Flux densities ($F_\nu$, from VLBI model
  fitting and VLA pulse profiles), FWHM sizes, and ICRF positions for detected sources.}
\begin{tabular}{lcllclclc}
\hline
 Source      &  $\nu_{\rm obs}$ (GHz) & RA & DEC & $F_\nu$ (mJy) & $1
                                                                    \sigma
                                                                   $
                                                                    range
                                                                    &
                                                                      Size 
                                                                      (mas)
  & $1 \sigma$ range & VLA $F_\nu$ (mJy)\\
\hline
 J1745-2912 &  8.7  &      17:45:47.83043(23) &  -29:12:30.780(3) &
                                                                    0.037
                                                                  &
                                                                    [0.33,
                                                                    0.42]
                                                                    &
                                                                      1.7 & [0.3, 3.1]   & 0.056  \\
 J1746-2849  & 8.7 &        17:46:03.35736(12) &  -28:50:13.385(2) &
                                                                     0.025
                                                                  &
                                                                    [0.021, 0.029] &
                                                                     5.3
                                                                  &
                                                                    [2.6, 8.0]   & 0.011   \\
 J1746-2856  &  5.9 &      17:46:49.85480(6) &  -28:56:58.990(1) &
                                                                   0.067
                                                                  &
                                                                    [0.065, 0.071]
                                                                    &
                                                                   10.9
                                                                  &
                                                                    [9.7, 12.1] & 0.11 \\
 B1758-23    &  7.5  &      18:01:19.81488(60) &  -23:04:44.637(10) &
                                                                      0.41
                                                                  &
                                                                    [0.34,
                                                                    0.45] &
     12.5 & [10.3, 14.5]   &  \\
 B1809-176   &  4.5 &        18:12:15.85925(17) & -17:33:37.871(2) &
                                                                     0.15
                                                                  &
                                                                    [0.09,
                                                                    0.20]
                                                                    & 
    17.6 & [10.3, 24.7]    &  \\
 B1822-14    &  7.5   &    18:25:02.95832(1) & -14:46:53.3605(2) &
                                                                   0.27
                                                                  &
                                                                    [0.26, 0.28] &1.8 & [1.7, 2.0]  &    \\
\hline

\end{tabular}
\end{table*}

\begin{figure*}
\begin{center}
\begin{tabular}{lll}
\includegraphics[scale=0.47]{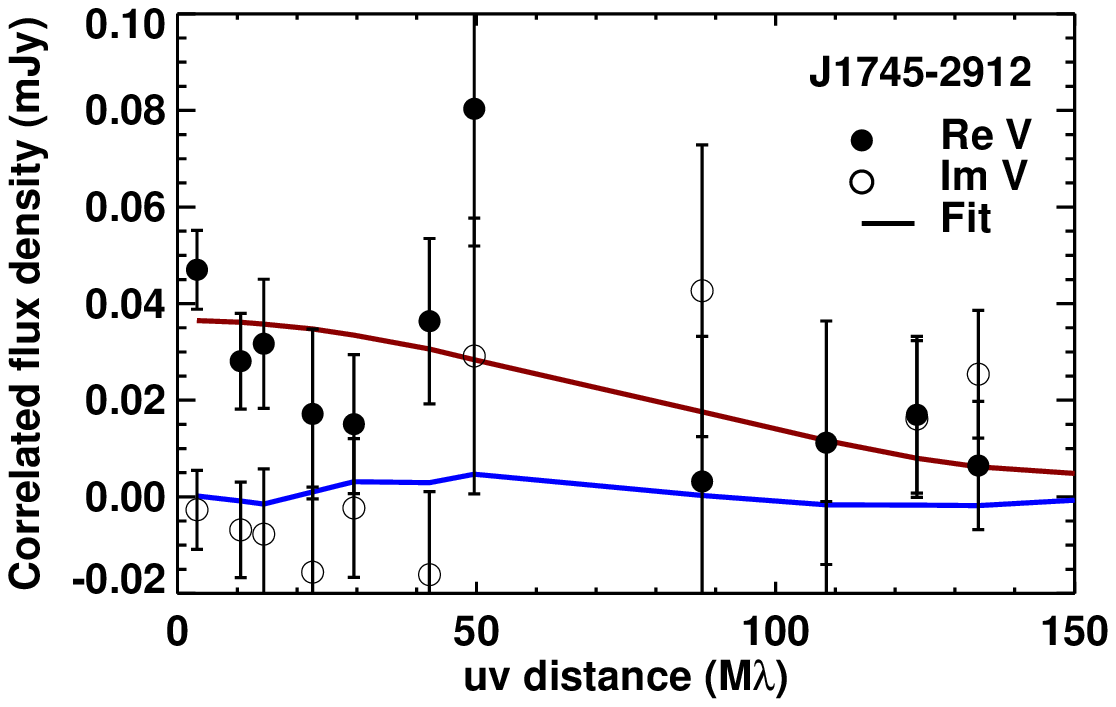}&
\includegraphics[scale=0.47]{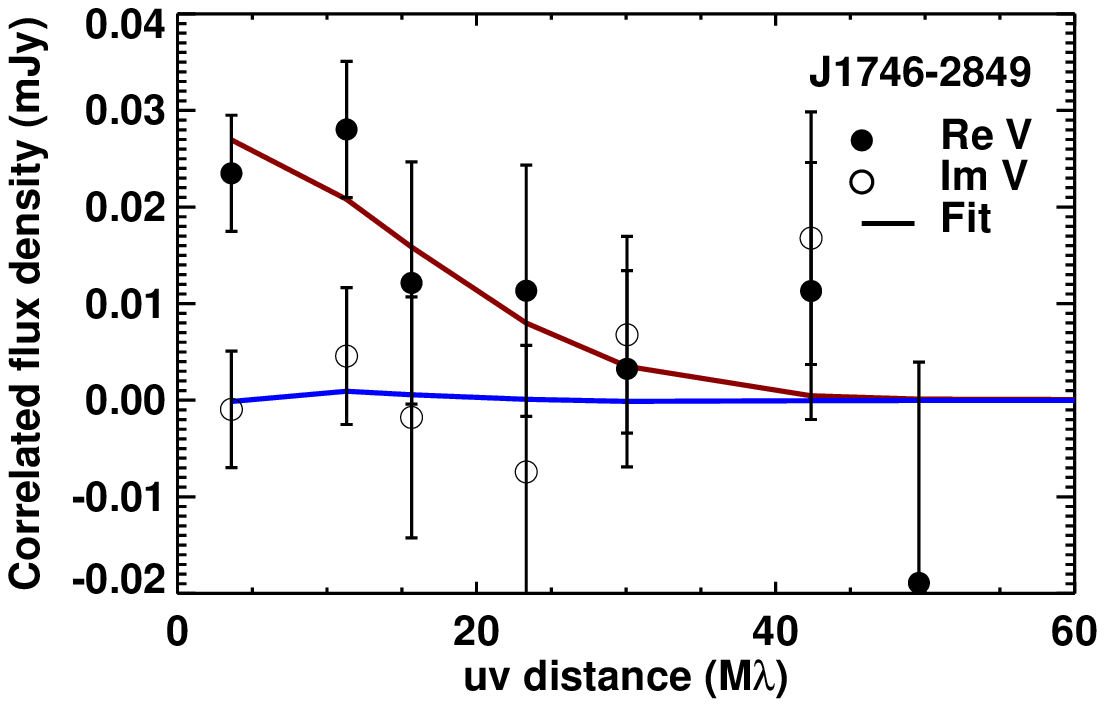}&
\includegraphics[scale=0.47]{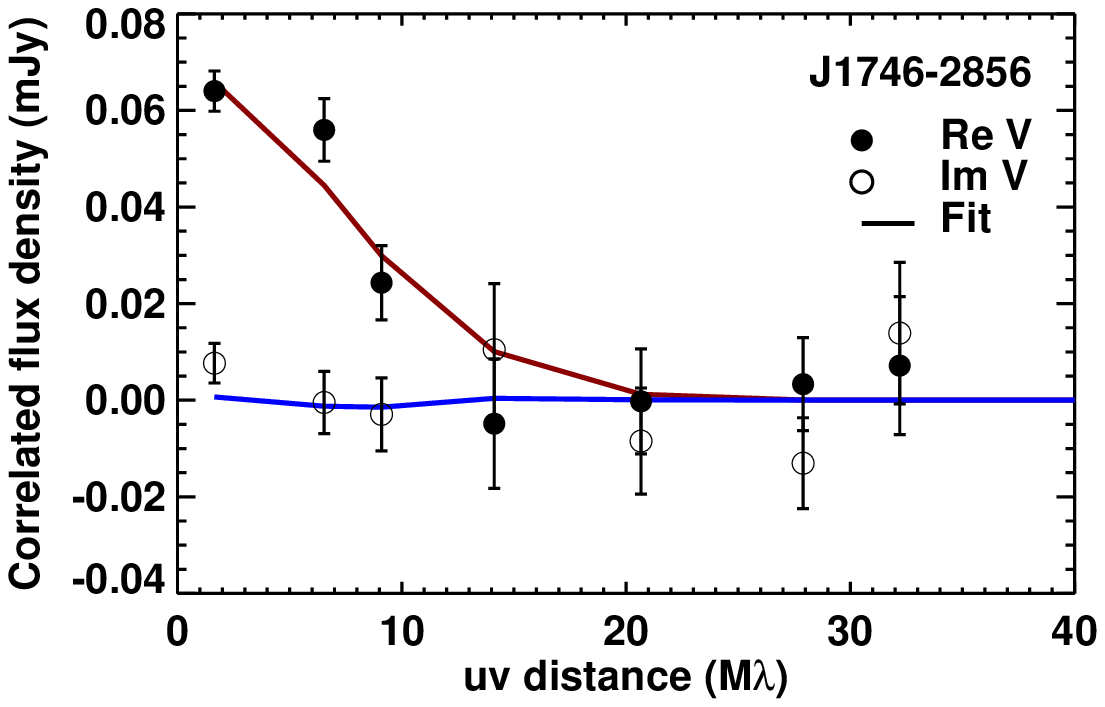}\\
\includegraphics[scale=0.47]{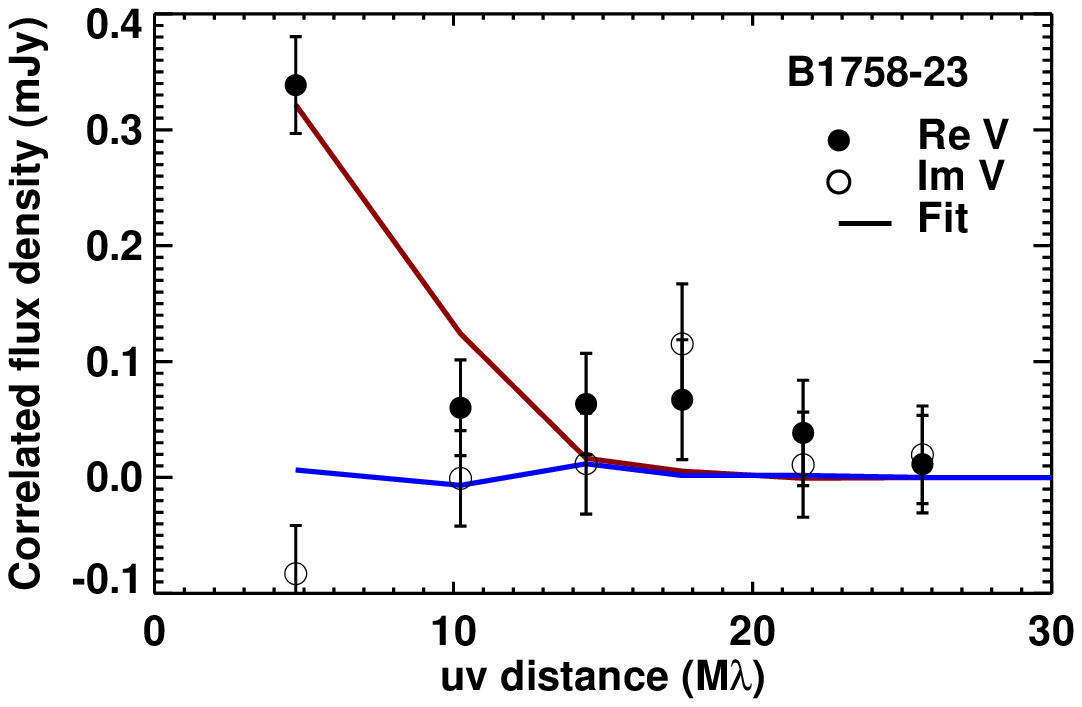}&
\includegraphics[scale=0.47]{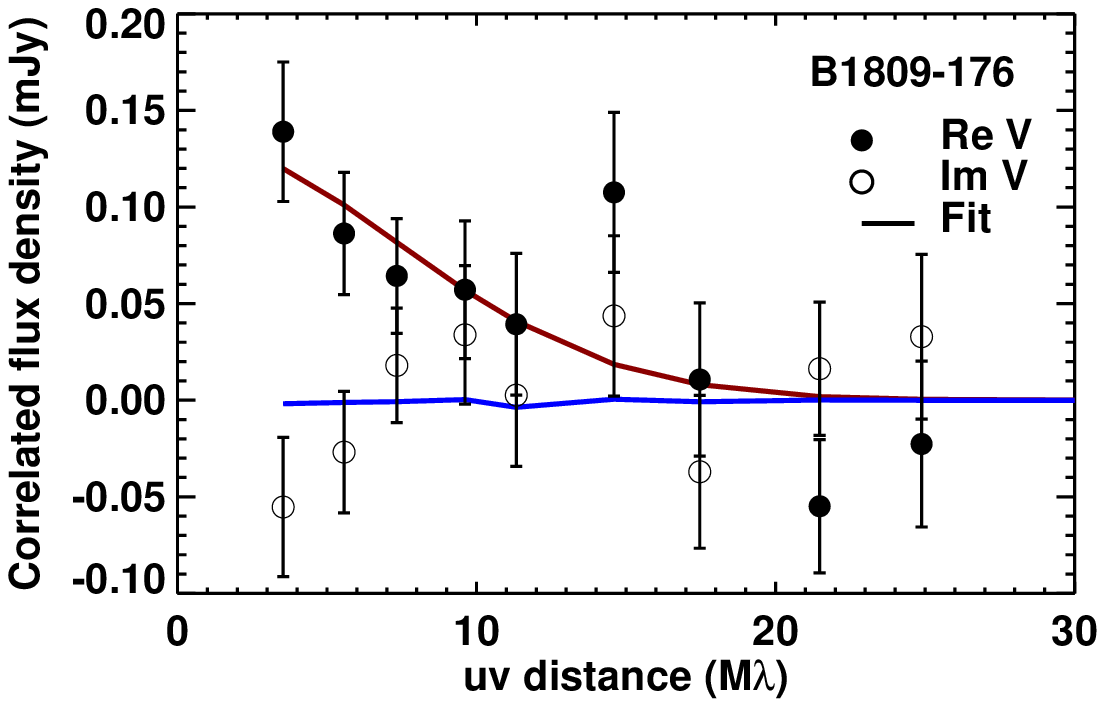}&
\includegraphics[scale=0.47]{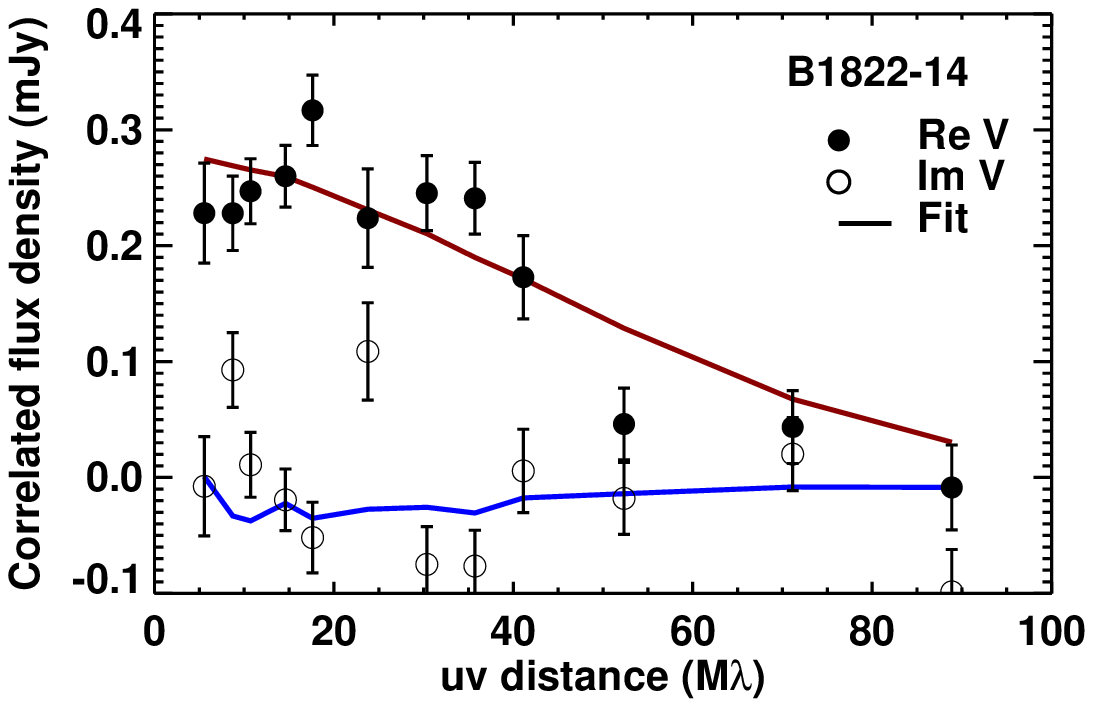}
\end{tabular}
\caption{\label{uvsizefits}Best fitting 1D offset Gaussian models (lines)
  compared to real (solid) and imaginary (open) visibilities
  for each source detected. For J1745$-$2912, we have shown the $8.7$ GHz
  data, as described in the text. The amplitude shown is the
  equivalent period-averaged flux density for the pulsar,
  i.e. correcting for the gate width.}
\end{center}
\end{figure*}

\subsection{PSR B1750$-$24}

We did not detect PSR B1750$-$24 in the gated image, despite a
predicted flux density and scatter-broadened image size similar to
that of PSR B1758$-$23, which was
detected in the same observation, and comparable image rms noise. Either the source is fainter
at $7.5$ GHz (e.g. because of a break in the spectrum) or it is more scatter-broadened than expected.

\subsection{PSR J1745$-$2912}

The GC pulsar J1745$-$2912 was found to have a small angular size at $8.7$ GHz (top right panel of Figure
\ref{images}), especially interesting since its temporal broadening was
found to be larger even than the GC magnetar
\citep{denevaetal2009detect}. To test the frequency-dependence of the
angular size, we re-observed J1745$-$2912 at $5.9$ GHz in the same
observation as J1746$-$2856. 

Unfortunately, the 5.9 GHz data were strongly affected by RFI, which limits their
reliability.  The folded data from the VLA tied-array beam shows
the pulsar signal with $F_\nu \simeq 0.1$ mJy, comparable to the flux
density at 8.7 GHz. However, in the gated image the brightest peak is seen several hundred milliarcseconds
(many synthesized beams) away from the source position at 8.7 GHz.
Moreover, this same peak is seen in the ungated image, and appears to
be largely generated by the shortest baseline (VLA to Pie Town), which
is likely the most RFI-prone.  If this baseline is flagged, no
significant source remains in the gated 5.9 GHz image.  Accordingly,
we make use only of the 8.7 GHz data for this pulsar; however, the
failure to detect the pulsar in the 5.9 GHz image given the clear
detection of pulsations in the tied-array beam is puzzling.

\begin{figure*}
\begin{center}
\begin{tabular}{l}
\includegraphics[scale=0.48]{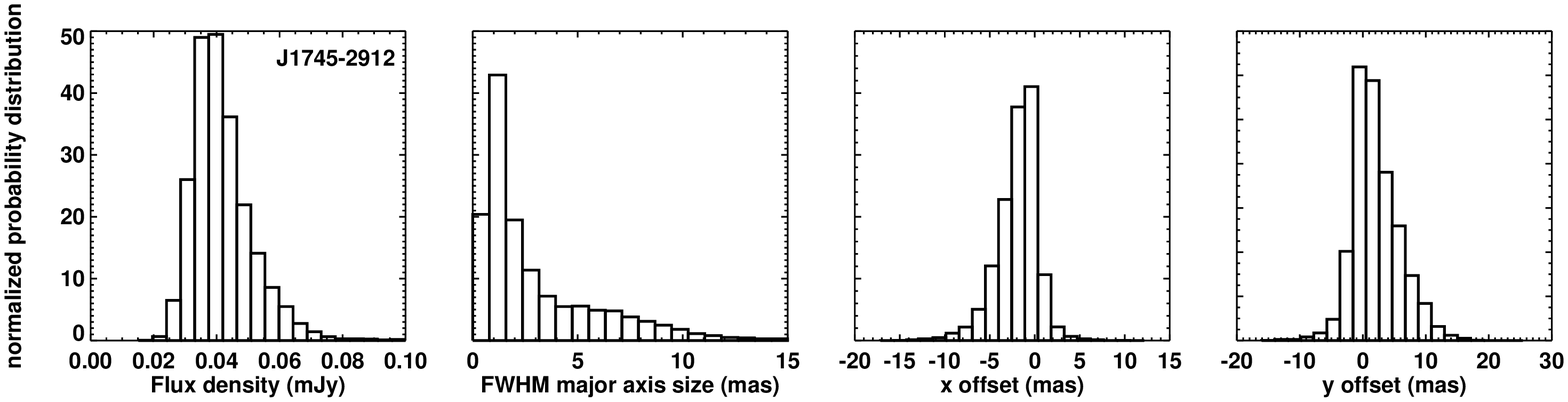}\\
\includegraphics[scale=0.48]{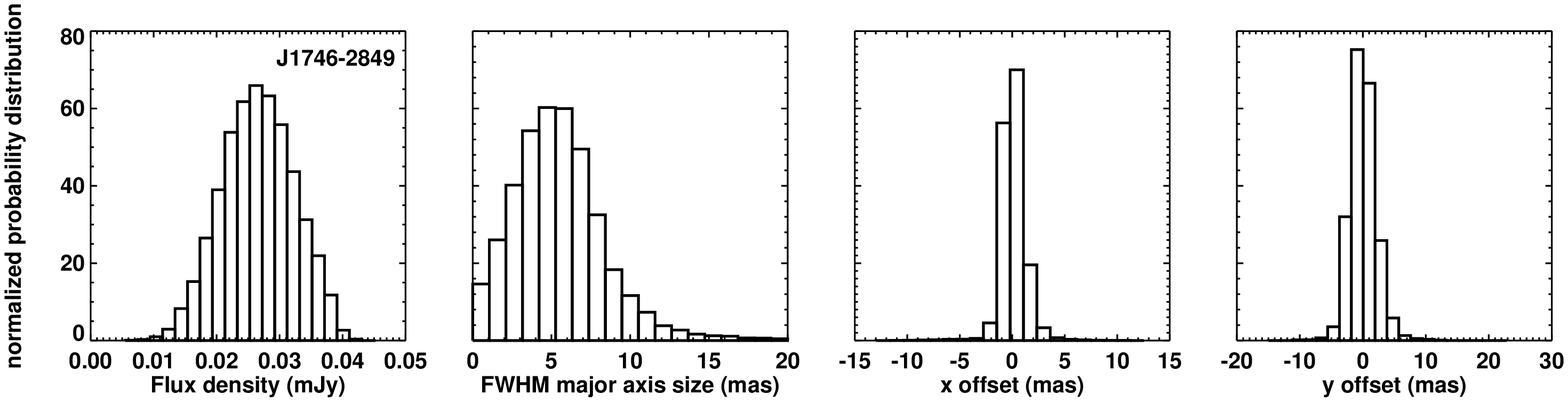}\\
\includegraphics[scale=0.48]{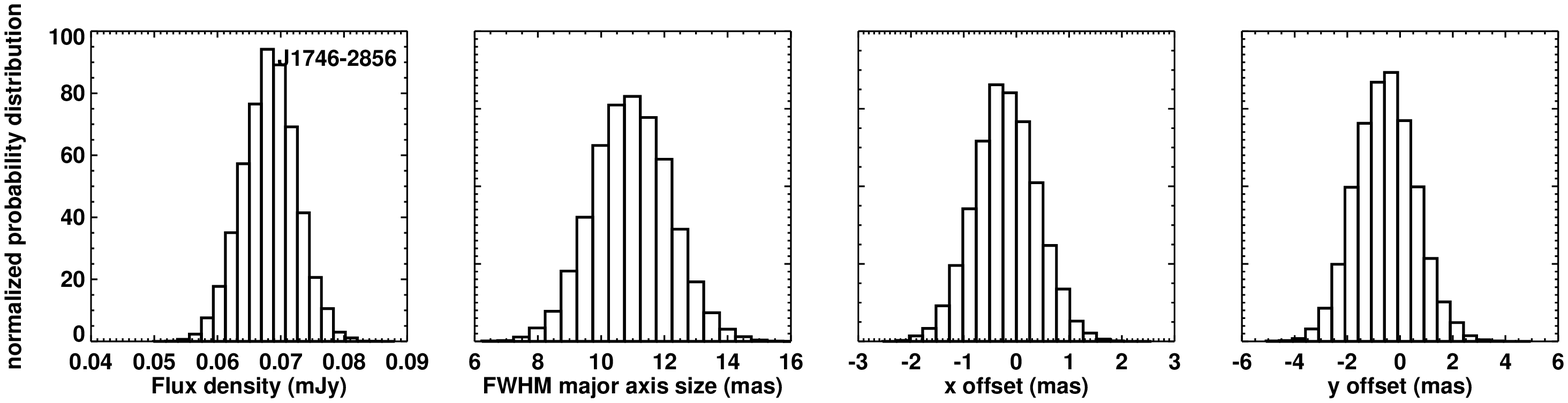}\\
\includegraphics[scale=0.48]{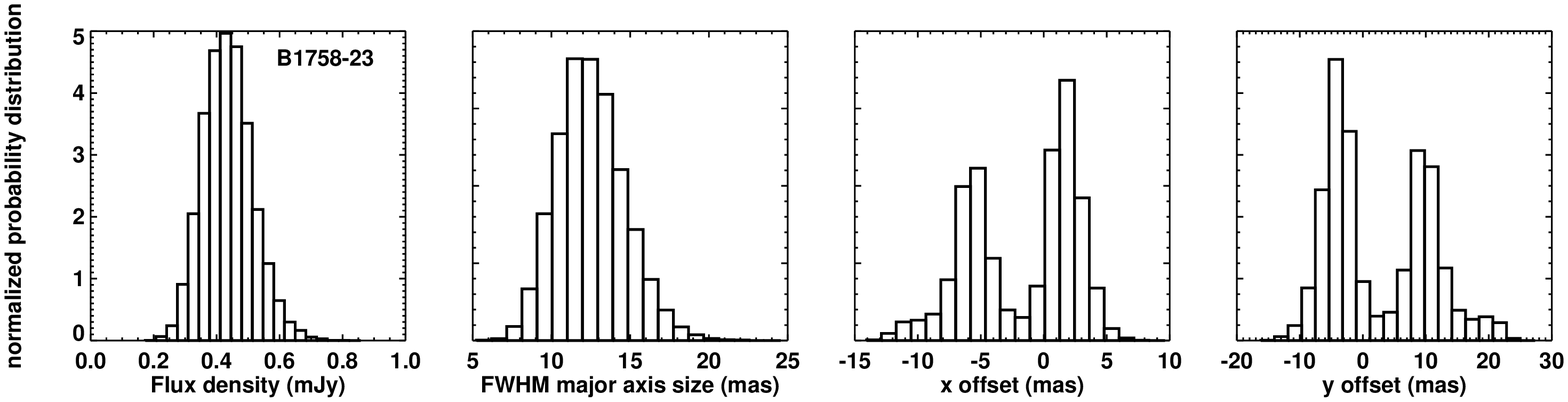}\\
\includegraphics[scale=0.48]{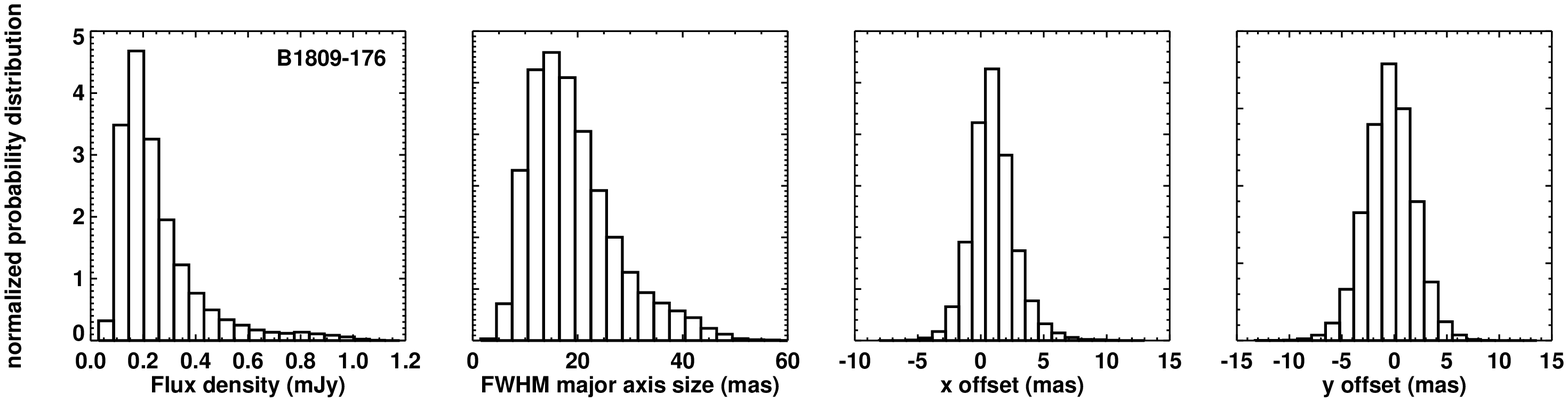}\\
\includegraphics[scale=0.48]{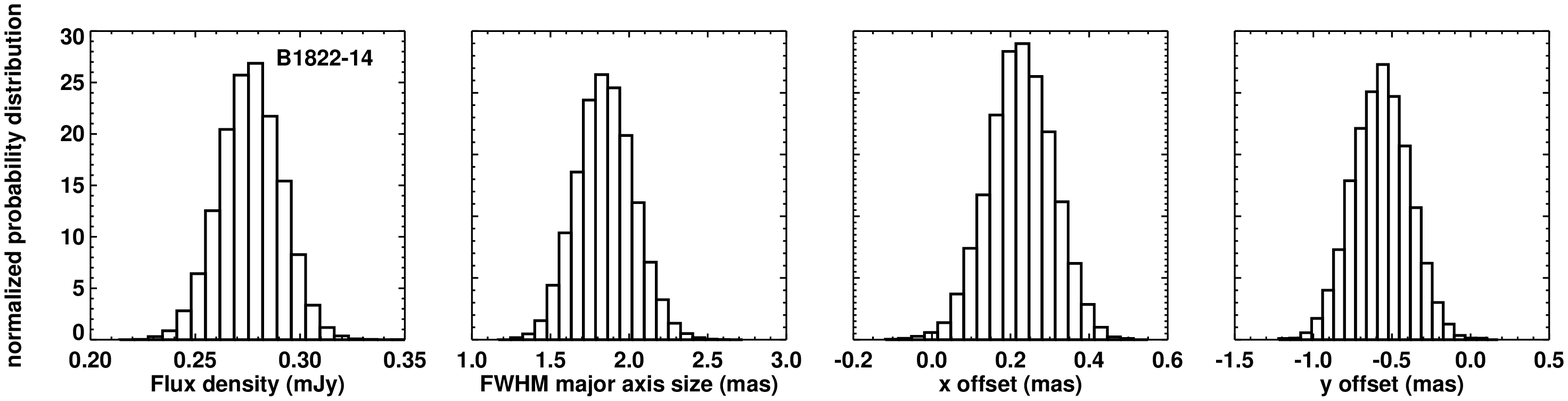}
\end{tabular}
\caption{\label{paramdists}Probability density as a function of total
  flux density, FWHM Gaussian size, and x and y offset from image
  centroid for all detected sources from fitting a symmetric Gaussian model to the scan-averaged
  calibrated visibilities. An extended source is detected in each case
  at a small offset from the image source position.}
\end{center}
\end{figure*}

\section{Size and position measurements}
\label{sec:size-posit-meas}

For the detected sources and using only the $8.7$ GHz data for
J1745$-$2912 (see above), we moved the phase centers of the visibilities
to the source positions found in the cleaned images, and measured
angular sizes and positions by fitting symmetric, offset Gaussian models to
the complex visibility data, averaged over scans (typically $\simeq
20$s). The parameter space over source flux density, angular FWHM
size, and (x,y) position offset was sampled using a Markov Chain Monte
Carlo (MCMC) algorithm as implemented in the publicly available \textsc{emcee} code
\citep{emcee}. The likelihood was calculated assuming uniform priors
on each parameter. A log(size) prior leads to
marginally smaller size estimates ($< 1 \sigma$) for the weakly
detected sources. We also tried asymmetric (2D) Gaussian models, since
the images seem to show asymmetric structure. However, asymmetry was
not significantly detected for any object ($< 2\sigma$). This is
probably a result of low signal-to-noise on individual baselines.

The best fitting models are
compared to scan-averaged and uv-binned data in Figure
\ref{uvsizefits}. The uv-binning is done for
presentation and was not included for fitting. The probability distributions over model parameters 
are shown in Figure~\ref{paramdists} and the best fitting flux
densities, source
positions, and FWHM angular sizes along with $1 \sigma$ confidence
intervals are listed in Table \ref{sizetable}. 

As expected, the model fits identify sources close to the
positions where they appear in our images in all cases. Source
extension is detected in
all cases, although with relatively low significance ($\simeq 90\%$)
for J1746$-$2849 and J1745$-$2912. For J1746$-$2849, this is due to the
faintness of the source. The size of J1745$-$2912 at $8.7$ GHz is much
smaller than the beam (e.g. Figure~\ref{images}) and so the source is only
partially resolved. Typical $1 \sigma$ uncertainties are $\simeq
10-50\%$. Residual phase errors likely
lead to systematic errors of comparable magnitude (\S~\ref{sec:syst-uncert}).

The flux densities from model fitting are also generally
compatible with (within a factor of 2 of) 
the expected values based on the known pulsar brightness and spectral
index values (Table~\ref{sources}). In particular, PSR B1758$-$23 must
have a relatively flat spectrum to be detected at $7.5$ GHz. We also 
confirm the steep spectrum of PSR J1746$-$2856. For the GC sources, we
can also compare to the independent VLA flux density estimates
(\S\ref{sec:vla-phased-array}). PSR J1746$-$2849 is found to be significantly brighter
in model fitting (factor of $\simeq 2.5$), while otherwise the
agreement is good within errors.

The image centroids (source positions) in most cases are constrained
to $\lesssim$ (2,4) mas. The best position constraint is for
B1822$-$14 ($0.1$,$0.2$) mas, where the detection significance is
very high and the source is compact. The precision is lower for more
extended sources ($1$,$2$) mas for J1746$-$2856, J1746$-$2849, and 
B1809$-$176), and lower still for J1745$-$2912 where the size is poorly
constrained ($2$,$4$) mas. For B1758$-$23, phase errors lead to an
elongated streak in the image and multiple solutions for the
source position reaching out to tens of mas offsets from the cleaned
image peak. Despite this issue the 1D source size remains well
constrained. The \emph{accuracy} of the source positions is limited to
$\gtrsim 1$ mas by the precision of the phase calibrator position in
the ICRF. The measured flux densities and angular sizes are consistent with the
images and with fits to uv-binned or time-averaged data.

All three GC pulsar locations are in good agreement with those from
recent ATCA observations \citep{schnitzeleretal2016}, with offsets
$\lesssim (0.020,0.2)$ arcsec. The offsets from
previous pulsar timing positions, both for GC and non-GC sources, are
larger: $\lesssim (0.3,0.6)$ arcsec, but generally in agreement within
errors. For B1809$-$176 the RA offset is $\simeq 4\sigma$ from the
pulsar timing position and for B1822$-$14 both offsets are $\simeq
2\sigma$. 

Our VLBA+VLA measurements used Sgr~A* as a secondary phase calibrator,
and so included the GC magnetar SGR J1745$-$2900 as well. Its angular
size is found to agree with previous measurements
\citep{boweretal2014,boweretal2015}. \citet{claussenetal2002}
previously constrained the size of
B1758$-$23 to be $< 0.5$ arcsec at $1$ GHz. Scaling our results to this
frequency assuming $\theta \propto \nu^{-2}$ gives $\theta \simeq 0.7
\pm 0.2$ arcsec, marginally compatible with their result. A flatter
scaling (see \S~\ref{sec:temp-broad-dist}) leads to better agreement.

\begin{table*}
\caption{Multi-frequency temporal broadening, $\tau (\nu_\tau)$, and distance ($D$)
  data used, and our extrapolation of the temporal broadening data to the observed
  frequency of the VLBI observations, $\tau (\nu_{\rm obs})$, using a
  spectral index $\tau \propto \nu^{-\alpha}$.\label{tautable}}
\begin{footnotesize}
\begin{center}
\begin{tabular}{ccccccccc}
        \hline
$\nu_\tau$ (GHz) & $\tau$ (ms) & $\alpha$ & $\nu_{\rm obs}$ (GHz) &
                                                         $\tau(\nu_{\rm
                                                         obs})$ (ms) &
                                                                       $D$
                                                                       (kpc)
  & $D_{\rm NE2001}$ & $D_{\rm YMW17}$ & Refs.\\
        \hline
     J1745$-$2912 & & & & & &\\
                     $3.1$   &  $25 \pm 3$   & $4 \pm 0.5$ &  8.7 &
                                                                    $0.4^{+0.3}_{-0.2}$
                                                                     &
                                                                       $8.3$
  & $15$ & $8.1$ &6,8,9,10,12\\
  J1746$-$2849 & & & & & &\\
1.5 & $266$ & $3.3 \pm 0.3$ & 8.7 & $0.9^{+0.7}_{-0.3}$ & $8.3$ & $30$
                     & $8.2$ &6,9,10,11,12\\
      2.0   &   $140$  & & &  & \\
     J1746$-$2856 & & & & & &\\
               1.4  &  $170 \pm 15$ & $3.07 \pm 0.14$ & 5.9 & $2.0 \pm
                                                              0.4$ &
                                                                     $8.3$
  & $8.4$ & $8.2$ &6,8,9,10,12\\
     3.1  &   $15 \pm 2$  &  &  &  & \\
      B1758$-$23 &  & & & & &\\
       1.275   &   $130.5 \pm 5.4$    & $3.5 \pm 0.2$ &   7.5 &  $0.27
                                                               \pm 0.10$ & $4
                                                                   \pm
                                                                   1$
  & $12$ & $6.5$
                                                             &
                                                               1-5,12\\
       1.374   &   $102.5 \pm 1.1$  &  & &  &  & \\
       1.400   &   $99 \pm 19$  &  &  & & &\\
       1.400   &   $111 \pm 19$  &   &  & & & \\
       1.421   &   $83.2 \pm 3.9$   &  &  & & & \\
       1.518   &   $74.3 \pm 1.3$   &   & & & & \\
       1.642   &   $51 \pm 10$  &   &  & & & \\
       1.642   &   $55 \pm 10$  &   &  & & & \\
       2.263   &   $17.9 \pm 0.8$  &  &  & & & \\
       2.600   &   $0.75 \pm 0.34$   &  & &  &  & \\
  2.700 & $8.6 \pm 1.7$ & & & & & \\
     4.850   &   $0.23 \pm 0.08$  &  & & &  & \\
       B1809$-$176 & & & & & & \\
                       $1$   &  $5.89 \pm 20\%$  & $4 \pm 0.5$  &    4.5 &  $0.14^{+0.14}_{-0.07}$
                                     &$6.3\pm 0.6$ & $6.3$ & $4.5$ & 6,7,12\\
      B1822$-$14 & & & & &\\
           $0.610$   & $143 \pm 31$  & $3.8 \pm 0.4$  & 7.5 & $0.010^{+0.007}_{-0.005}$ & $5.5
                                                                \pm
                                                                                          0.5$
  & $5.5$ & $4.5$
                                                             & 3,6,12\\
     1.060   &   $15.1 \pm 2.0$  &  &   &  & \\
     1.400   &   $6.1 \pm 1.2$  &   &  &  & \\
      1.642   &   $3.7 \pm 1.5$  &   &  &  & \\
	\hline
%\begin{tablenotes}
%\end{tablenotes}
\end{tabular}
\\
\end{center}
Reference key: (1) \citet{manchesteretal1985} (2)
\citet{loehmeretal2001} (3) \citet{lewandowskietal2013} (4)
\citet{lewandowskietal2015} (5) \citet{verbiestetal2012} (6)
\citet{cordeslazio2002} (7)
\citet{manchesteretal2005} (8)
\citet{johnstonetal2006} (9) \citet{chatzopoulosetal2015} (10)
\citet{gillessenetal2017} (11)
\citet{denevaetal2009detect} (12) \citet{yaoetal2017}
\end{footnotesize}
\end{table*}

\begin{table*}
\caption{Screen distance $D_s$ calculation results from our
  observations compared to those of J1745$-$2900.\label{dstable}}
%\begin{small}
\begin{center}
\begin{tabular}{lcccccccc}
        \hline
Name & $\nu$ (GHz) & $\tau$ (ms) & $D$ (kpc) & $\theta$ (mas) & $D_s$ (kpc)\\
        \hline
     J1745$-$2912 &     8.7 & $0.4^{+0.3}_{-0.2}$ &$    8.3$ &$
                                                             1.7\pm
                                                             1.4$ &$
                                                                    8.0
                                                                    \pm
                                                                    0.3$ &\\
     J1746$-$2849 &     8.7 &      $0.9^{+0.7}_{-0.3}$ & $    8.3$
                                             &$5.3\pm 2.7$ &$7.4 \pm 0.7$
                &\\
     J1746$-$2856 &     5.9 &     $2.0 \pm 0.4$ &$    8.3$ &$   10.9\pm
                                                           1.2$ &$
                                                                  6.9
                                                                  \pm 0.3$ &\\
       B1758$-$23 &     7.5 & $0.27 \pm 0.10$ &$    4 \pm    1$ &$   12.4\pm    2.1$ &$    2.0\pm    0.5$ &\\
      B1809$-$176 &     4.5 & $0.14^{+0.14}_{-0.07}$ &$    6.3\pm    0.6$ &$   17.5\pm    7.2$ &$    1.0\pm    0.7$ &\\
       B1822$-$14 &     7.5 & $0.010^{+0.007}_{-0.005}$ &$    5.5\pm
                                                        0.5$ &$
                                                               1.85\pm
                                                               0.15$
                                                              &$
                                                                2.9\pm
                                                                0.8$
                &\\
  \hline
  J1745-2900 & 8.7 & $0.35 \pm 0.10$ (1) & $8.3$ & $12.5 \pm 1.2$ (2) & $3.1
                                                                \pm
                                                                0.6$ & \\
	\hline
\end{tabular}
\end{center}
(1) \citet{spitleretal2014} (2) \citet{boweretal2014}
%\end{small}
\end{table*}

\begin{figure}
\begin{center}
\includegraphics[scale=0.7]{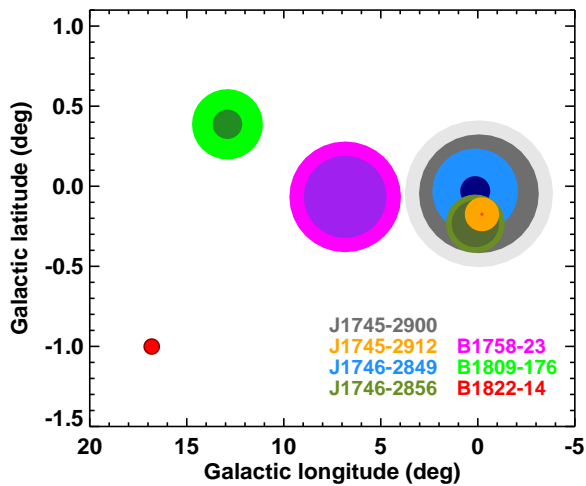}
\caption{\label{sizemap}Measured $\pm 1\sigma$ angular size ranges
  (radii of the inner and outer circles) vs. Galactic
  coordinates ($l$,$b$), scaled to the angular broadening of the
  GC magnetar SGR J1745$-$2900 \citep[gray,][]{boweretal2014}. The
  other GC sources are significantly smaller in angular extent than
  J1745$-$2900 and Sgr~A*, while B1758$-$23 is comparable in angular size.}
\end{center}
\end{figure}

\begin{figure*}
\begin{center}
\begin{tabular}{ll}
\includegraphics[scale=0.72]{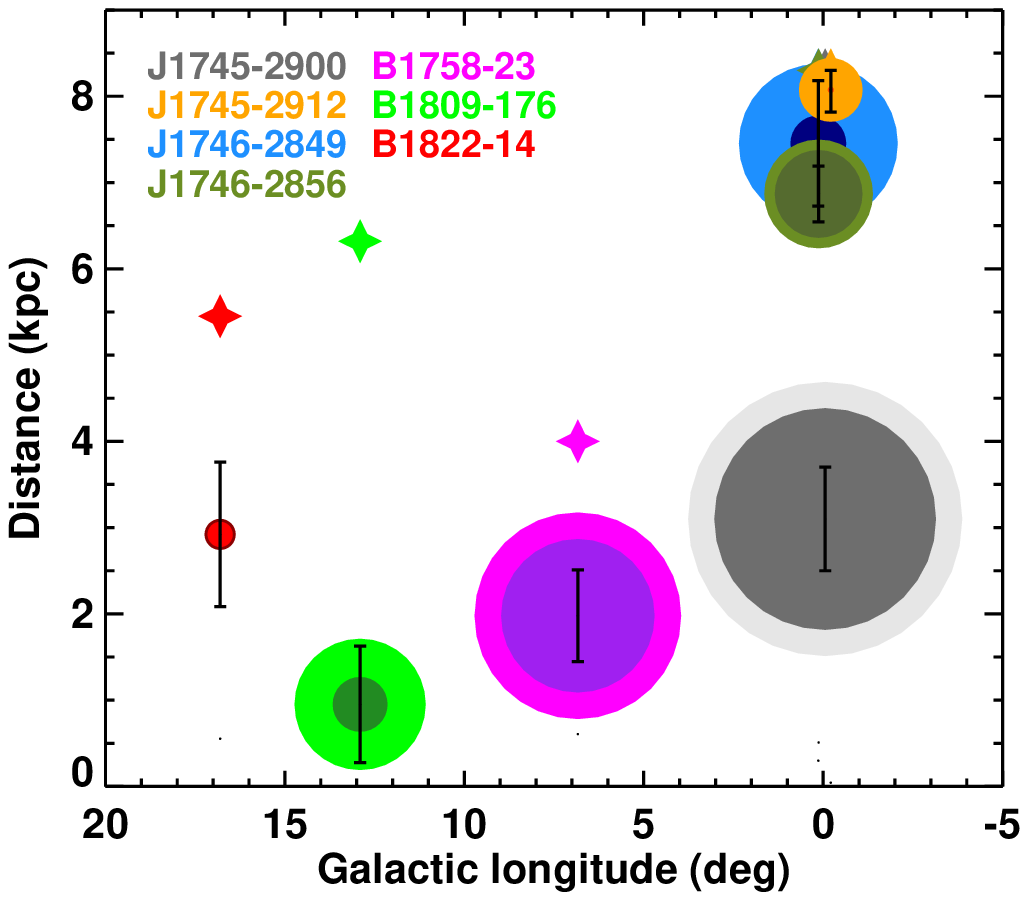}&
\includegraphics[scale=0.72]{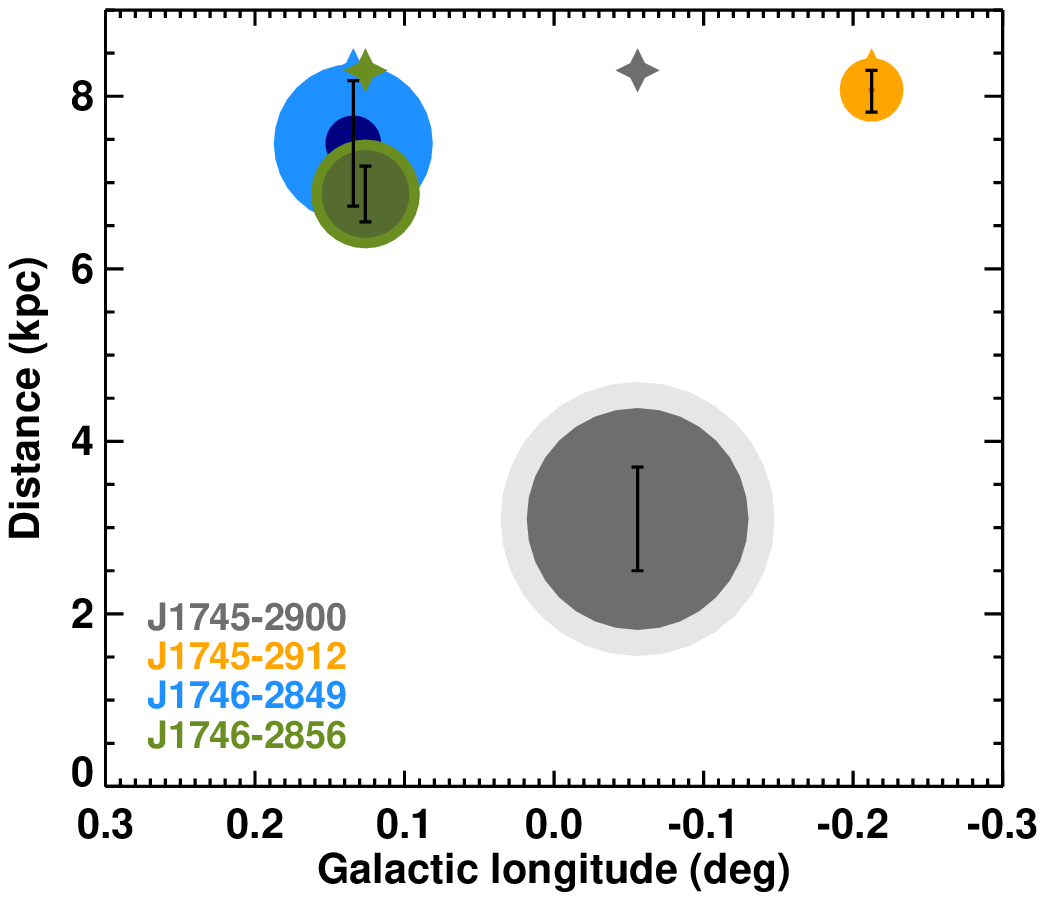}
\end{tabular}
\caption{\label{locmap}Measured $\pm 1\sigma$ angular size ranges
  (radii of the inner and outer circles, scaled to the size of the GC
  magnetar in gray) vs. Galactic longitude and
  distance from Earth for all detected sources (left) and zooming in
  on the GC sources (right). The stars show the location of the
  pulsars, while the circles are placed at the scattering location
  $D_s = D - \Delta$ found from the
  combined angular and temporal broadening 
  (equation \ref{eq:1}). The sources near the GC are
  assumed to be located at the distance of Sgr~A*
  \citep{chatzopoulosetal2015}, while the DM 
  distance is used for the remaining sources. The location of the
  scattering towards all 3 non-GC sources detected is consistent with
  a nearby spiral arm. The scattering towards
  the other GC sources is inferred to occur much closer to the GC than
in the case of Sgr~A* and SGR J1745$-$2900. The very small size of
J1745$-$2912 implies $\Delta \lesssim 700$ pc, direct evidence for a
strong scattering medium in the GC.}
\end{center}
\end{figure*}

\begin{figure}
\begin{center}
\includegraphics[scale=0.85]{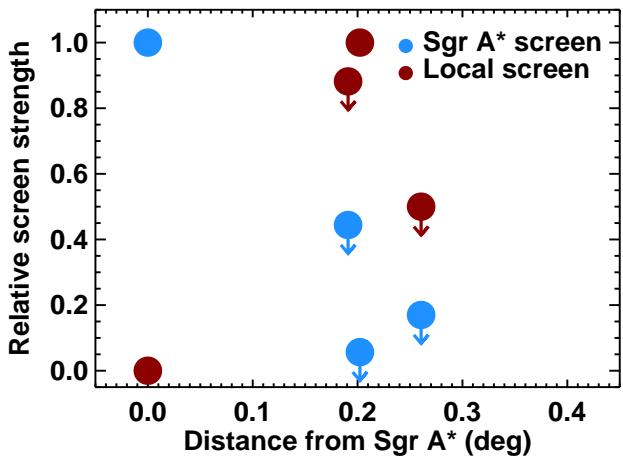}
\caption{\label{gcscreen}Maximum fraction of the temporal
  broadening of the GC pulsars (blue points) that could be
  produced at the scattering location of J1745$-$2900, $\Delta \simeq 5$
  kpc from the GC \citep{spitleretal2014,boweretal2014}, as a 
function of their angular separation. The scattering source for the
magnetar and Sgr~A* weakens or disappears on scales 
$\lesssim 0.2$ deg. The bulk of the temporal broadening for these
pulsars must have a different physical origin, closer to or residing
within the GC. The maximum relative strength of a possible screen
local to the GC, suggested by the small tentative size measurement of
J1745$-$2912, is shown as the red points. The two screen model (red
and blue points) constitutes the minimal assumption needed to explain
all GC pulsar observations. The upper limits take into account models
with additional components.}
\end{center}
\end{figure}

\begin{figure*}
\begin{center}
\begin{tabular}{ll}
\includegraphics[scale=0.7]{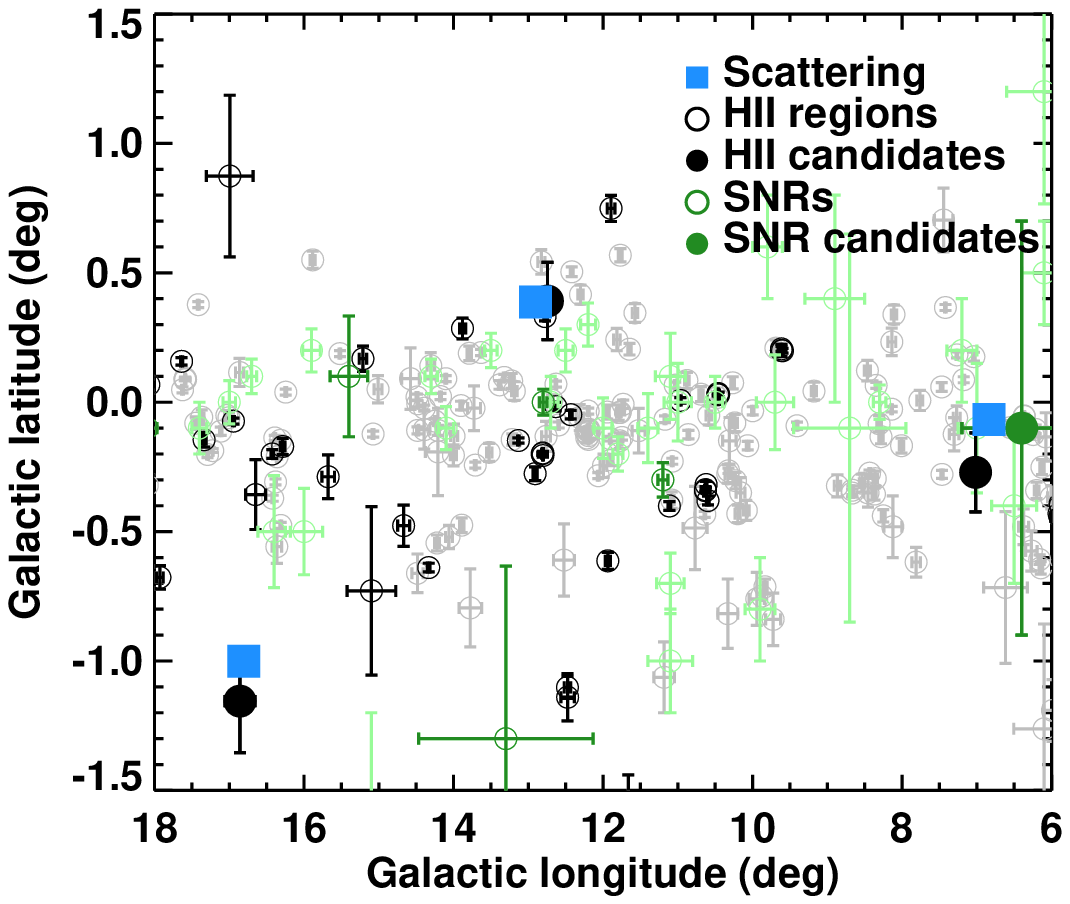}&
\includegraphics[scale=0.7]{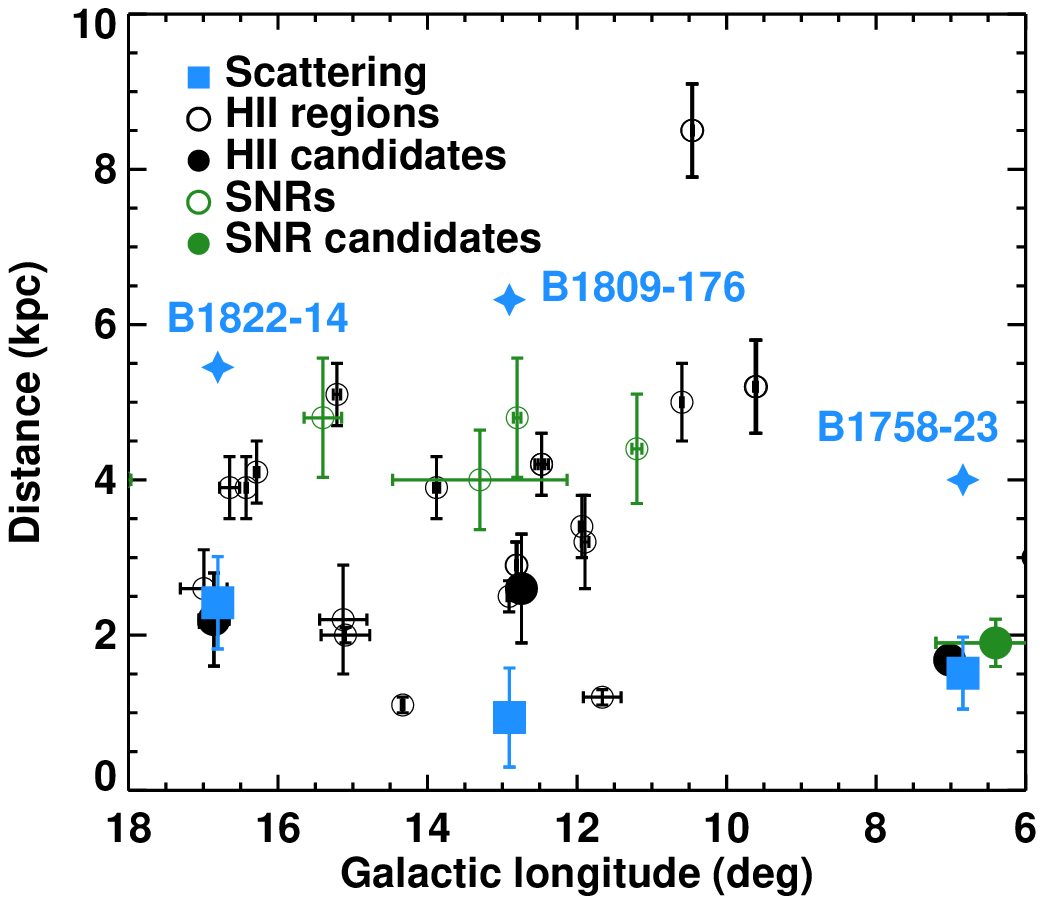}
\end{tabular}
\caption{\label{candidatemap}Scattering locations towards the non-GC
  pulsars in our sample (light blue squares) and the pulsars
  themselves (light blue stars, right panel) compared with the positions of known HII
  regions \citep[open black points,][]{andersonetal2014} and supernova remnants \citep[open green
  points,][]{green2014} in Galactic longitude and latitude (left) and Galactic
  longitude vs. distance (right). The error bars in Galactic
  coordinates correspond to the measured sizes of the objects, while
  the error bar in the distance is its uncertainty. In the right panel
  we only show objects for which distances are given in the catalogs
  (darker points in the left panel). We identify
  candidates (solid circles) as sources overlapping with the (l,b) position of our
  pulsars. In all cases, these sources have distances commensurate
  with our inferred distance to the scattering medium in front of the
  pulsars. For all three objects, an HII region has the right distance
to produce the observed scattering. For B1758$-$23, the supernova
remnant W28 is also at the inferred scattering location (but closer to
Earth than PSR B1758$-$23, see figure \ref{b1758dist}).}
\end{center}
\end{figure*}

\begin{figure}
\begin{center}
\includegraphics[scale=0.7]{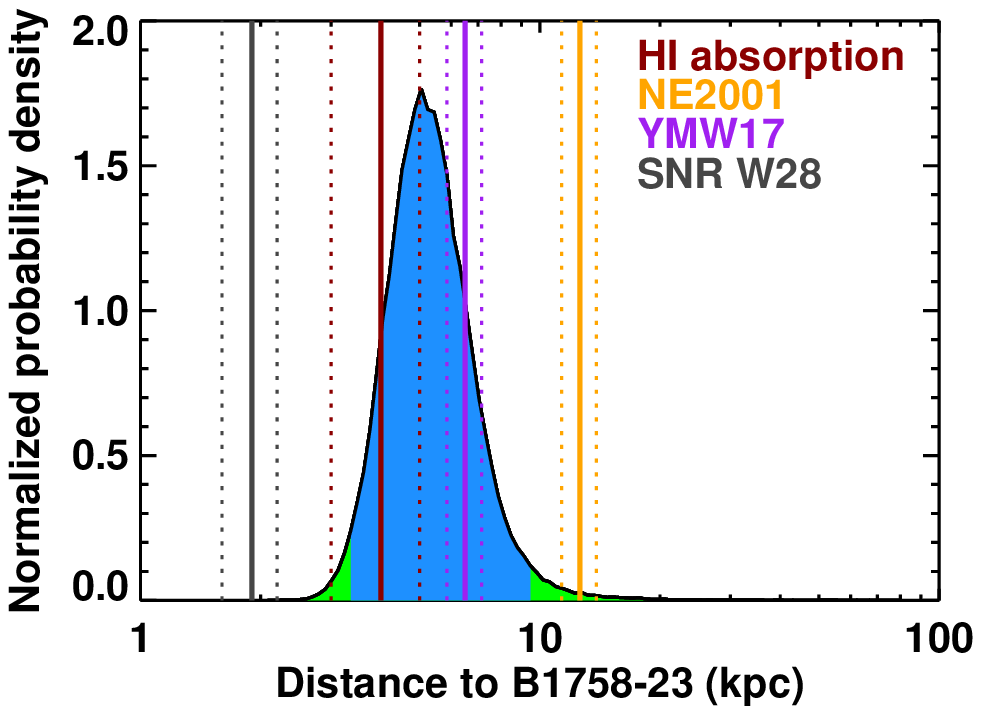}
\caption{\label{b1758dist} Probability distribution for the distance
  to the pulsar B1758$-$23 measured from combining its temporal and
  angular broadening (Table \ref{tautable}) with the angular broadening
  of the nearby extragalactic background source J1801-231
  \citep{claussenetal2002}, assuming the two sources are behind a
  common, thin scattering medium. The resulting distance estimate
  (blue shaded region shows $2\sigma$ confidence interval) is
  consistent with the recent measurement of $4 \pm 1$ 
  kpc from HI observations \citep{verbiestetal2012} and the recent DM
  distance from \citet{yaoetal2017}, but smaller than
  the prediction of the NE2001 model \citep{cordeslazio2002}. This
  distance is also incompatible with an association of the 
  pulsar with the supernova remnant W28 at a distance $1.9 \pm 0.3$
  kpc \citep{velazquezetal2002}, although this SNR could contribute
  to its scatter broadening (Figure~\ref{candidatemap}).}
\end{center}
\end{figure}

\section{Mapping strong interstellar scattering}
\label{sec:mapp-strong-interst}

Figure~\ref{sizemap} shows the angular sizes of the sources measured
here as a function of Galactic coordinates, scaled to the size of SGR
J1745$-$2900 at the observing frequency (equivalent to assuming $\theta \propto
\nu^{-2}$). Sgr A* and SGR J1745$-$2900 have been found to have
the same scatter-broadened image in size, position angle, and
frequency-dependence \citep{boweretal2014}. We show the scattering
properties of SGR J1745$-$2900 as a reference in what follows,
assuming that they are identical to those of Sgr A*.

The sources in the GC are
all found to be smaller in angular size than SGR J1745$-$2900. J1746$-$2849 and J1746$-$2856 are a factor $\simeq 2-3$ smaller, while
the very compact $8.7$ GHz size of J1745$-$2912 is several times smaller. These variations in angular size are comparable to
those among the known OH/IR masers \citep{vanlangeveldeetal1992,frailetal1994,yusefzadehetal1999} and
extragalactic background sources \citep{lazioetal1999,boweretal2001} at similar
separations. Outside of the GC, the pulsar B1758$-$23 has a large angular size,
comparable to that of Sgr~A*, despite a smaller $\tau$ and DM by
factors $\simeq 2$. The other sources are significantly less
scattered. B1809$-$176 still shows significant angular broadening, while
B1822$-$14 is compact. 

In the following, we combine these new angular broadening measurements with 
distance and temporal broadening values from the literature to locate
the scattering along the line of sight. We then identify candidate
origins for the scattering by comparing these locations 
with those of known HII regions and supernova remnants.

\subsection{Temporal broadening and distance data}\label{sec:temp-broad-dist}

We use existing data for the temporal broadening and distances to the
pulsars in our sample (Table \ref{tautable}). The literature data come first from
the ATNF catalog \citep{manchesteretal2005} and references within, and
further include more recent, multi-frequency measurements
\citep{lewandowskietal2013,lewandowskietal2015}. The GC pulsars are
assumed to be located at the distance of the GC, which we fix at $8.3$ kpc
\citep{reidetal2014,chatzopoulosetal2015,gillessenetal2017}. The
pulsar B1758-23 has a recent distance measurement of $4 \pm 1$ kpc
from HI absorption \citep{verbiestetal2012}. For PSRs B1809-176 and B1822-14
we use DM distances from the NE2001 model
\citep{cordeslazio2002}. Distances predicted using Galactic electron
density distribution models depend on the model employed; using the
recent YMW17 model \citep{yaoetal2017} in place of NE2001 predicts
significantly smaller distances for these two pulsars, which we find
to be more consistent with our scattering data.  However, in estimating
screen locations from our data we find insignificant ($< 0.3$ kpc)
differences from the choice of DM distance.

From the literature data, we estimate $\tau$ at the observing frequencies used
for our size estimates. This involves extrapolation: $\tau$ is difficult to
measure at high frequencies like those used for the VLBA+VLA observations,
which are required in order to match the array resolution to the large
image sizes. 

We extrapolate $\tau$ to the VLBA+VLA observed frequency using a
spectral index $\tau \propto \nu^{-\alpha}$. For a single, infinitely
extended, thin scattering screen this value is $\alpha \ge 4$, where
$\alpha = 4.4$ for Kolmogorov turbulence and $\alpha = 4$ for a finite
turbulent inner scale \citep{goodmannarayan1985}. When these model
assumptions break down, the frequency scaling is flatter
\citep{cordeslazio2001}, as seen for many high DM pulsars
\citep[average $\alpha \simeq 3.5$,][]{loehmeretal2001} like those imaged
here. When multiple measurements of $\tau$ are available, we use the
measured spectral index and account for the extrapolation error in our
final estimate of $\tau$ at our observing frequency. When only a
single value is available (PSR J1745$-$2912\footnote{We tried to
  measure $\tau$ at $5.6$ GHz from the tied-array VLA data. A value 
  of $\simeq 2-3$ ms is compatible with the data, but the result
  depends on the assumed intrinsic profile. A value $> 3$ ms seems unlikely, implying a slope
  $\alpha \gtrsim 3.5$.} and PSR B1809$-$176), we
assume $\alpha=4 \pm 0.5$ to extrapolate. The values of $\alpha$
assumed, their errors, and resulting $\tau$ estimates and errors are
shown in Table \ref{tautable}. For J1746$-$2849, the published values
\citep{denevaetal2009detect} show a very flat slope $\alpha \approx
2.2$. An upper limit on $\tau \lesssim 5$ ms at $5$ GHz comes from the observed
pulse width. Adding this limit leads to an estimate of $\alpha = 3.3 \pm 0.3$. For B1758$-$23, \citet{lewandowskietal2013} reported a steep
dependence $\alpha = 4.92$ using a mix of low- and high-frequency data. \citet{lewandowskietal2015} removed the high-frequency data and
found a shallower slope $\alpha = 3.62$. We use the latter measurement,
which is consistent with the result of \citet{loehmeretal2001}, and note that $\tau$ could be smaller if the high-frequency data are
more accurate. We further assume $10\%$ DM distance uncertainties
\citep{cordeslazio2002}, but they do not strongly affect the
results. Section \ref{sec:syst-uncert} includes additional discussion
of the systematic errors from extrapolating $\tau$ and using DM distances.

\subsection{Locating scattering screens}
\label{sec:locat-scatt-scre}

Given angular and temporal broadening measurements at the same
frequency to the same source,
the single thin screen scattering model gives a location for the
scattering medium of \citep{cordeslazio1997},

\begin{equation}\label{eq:1}
\frac{\Delta}{D} = \left(1+\frac{8 c \tau \ln 2}{D \theta^2}\right)^{-1},
\end{equation}

\noindent where $\Delta$ is the distance from the source to the
scattering medium, $D$ is the source distance, $\tau$ is the pulse
 broadening decay constant $\propto e^{-t / \tau}$, and $\theta$
is the angular broadening in terms of FWHM Gaussian image size. Using existing
$\tau$ measurements and distance estimates (above and Table
\ref{tautable}), we infer scattering locations $D_s = D-\Delta$ for
all objects in our sample.

To find median values for $D_s$ and its uncertainty for each pulsar,
we draw random Gaussian samples for $\tau$, $\theta$ (using our
measured $1\sigma$ errors from \S\ref{sec:size-posit-meas}), and $D$, calculate $D_s$ for each sample, and
measure $1\sigma$ confidence intervals based on their 
distributions. The resulting $D_s$ values are listed in Table
\ref{dstable}. The values from previous work for the GC magnetar are
also there, where we have extrapolated $\tau$ as above. Using our
method we find $D_s = 3.1 \pm 0.6$ kpc, compared to $D_s = 2.6 \pm
0.3$ kpc from \citet{boweretal2014}. The values are consistent within
$1 \sigma$. Ours is slightly larger and with larger uncertainty due to
extrapolating $\tau$ with $\alpha = 3.8 \pm 0.2$
\cite{spitleretal2014} rather than $\alpha = 4$ in \citet{boweretal2014}. 

Figure~\ref{locmap} again shows the measured source sizes, but now as
a function of Galactic longitude and line of sight distance. The vertical error bar
shows the scattering location $D_s$ and its uncertainty. The
measurement for SGR J1745$-$2900 \citep{boweretal2014} is closer to Earth than the inferred scattering locations
for the other GC pulsars. This is due to their relatively small
angular sizes and large $\tau$ values compared to those of the
magnetar. The very compact $8.7$ GHz size and large $\tau$ for
J1745$-$2912 would place its scattering local to the Galactic center:
$\Delta < 700$ pc. This is tentative evidence for strong scattering
from the hot, dense medium within the GC itself. The other GC pulsars have
scattering media $\simeq 7$ kpc from Earth. That location could
either arise from a single scattering origin at that location, or from a combination of distant scattering similar
to the magnetar and local scattering as seen for J1745$-$2912.

The non-GC pulsars all have $D_s \simeq 1-2$ kpc, consistent with
locations in the Carina-Sagittarius or Scutum spiral arm and similar to or
closer than the scattering medium producing the image of the GC
magnetar. The large size of B1809$-$176 implies a scattering location
$\simeq 1$ kpc from Earth. As discussed below, the nearest candidate
sources are closer to $\simeq 2.5$ kpc. The size could be
overestimated, the $\tau$ value could be underestimated (see also
\S \ref{sec:syst-uncert}), or the scattering could have some other
physical origin.

\subsection{Multiple scattering origins towards the GC}\label{sec:mult-scatt-orig}

The sizes of scatter-broadened maser and extragalactic background
sources within $\simeq 0.5$ deg of Sgr~A* have long been known to vary
by factors of several
\citep[e.g.,][]{vanlangeveldeetal1992,frailetal1994,yusefzadehetal1999,lazioetal1999,boweretal2001,pynzar2015}.
Our finding of small angular sizes for GC pulsars with large degrees of
temporal broadening demonstrates that these variations are not only
the result of varying strength in a single scattering medium. Instead,
they require (at least) a second physical component to the scattering. The small size found for
J1745$-$2912, if it holds, suggests this component could be located
within the GC itself ($\Delta < 700$~pc). The existence of such a component has long been
suggested \citep[e.g.,][]{cordeslazio1997}, but with $\tau$ a
factor $\sim 10^{2-3}$ higher than that of the known GC pulsars. Our
sample selection of known pulsars detected at GHz 
frequencies means that none of the identified scattering screens can be
responsible for obscuring long period pulsars in the GC. 

Phase-resolved angular broadening of the pulses of J1745$-$2900
shows that the scattering appears to be dominated by a single thin
scattering screen \citep{wucknitz2015}. In calculating screen 
  locations ($D_s$), we have assumed this to be true separately for each of the other pulsars
as well. Instead we now consider the minimal model needed to explain
all of the GC pulsar scattering measurements. The model consists of two scattering origins, one
local to the GC ($\Delta < 700$~pc to explain the small size of
J1745$-$2912) and one at $\Delta \simeq 5$ kpc as inferred for the
magnetar. For simplicity we assume that the local GC screen only contributes to $\tau$
and not $\theta$, while the other contributes to both as described by
equation \eqref{eq:1}. In this scenario, we calculate the 
contribution to the $\tau$ of our GC pulsars from the distant screen
and from the local GC screen in order to produce
their observed angular broadening. These contributions are shown in
Figure~\ref{gcscreen} (red and blue filled circles), scaled to the
$1.3$s $\tau$ measurement for J1745$-$2900 
\citep{spitleretal2014} and to the $\tau \simeq 2.3$s at 1~GHz of
J1745$-$2912 \citep{denevaetal2009detect}. By definition the magnetar
and J1745$-$2912 only have contributions from the distant and GC
screens respectively. This minimal model sets robust upper limits on
the contributions of the two components to the
angular broadening of each pulsar. The limits
(shown in the figure) generalize to include models with additional scattering
components.

The simple model robustly shows that the distant GC screen drops significantly in relative strength at the
location of the other GC pulsars, particularly for J1745$-$2912 because
of its very small size, but also (robustly) for J1746$-$2856 where the size is well
constrained. The constraint for J1746$-$2849 is weaker, because the size
is more weakly constrained. Conversely, the local GC screen does not
appear to contribute significantly at the location of J1745$-$2900,
since the scattering is well explained by a single screen, but could 
produce a large fraction of the observed temporal broadening for the
other GC pulsars given their scattering locations in or near the
GC. The observed maser sources
\citep[e.g.,][]{vanlangeveldeetal1992,frailetal1994} are heavily scatter-broadened out to scales of $\simeq 0.5$ deg,
larger than the scale over which the two components vary greatly in
strength. The physical medium responsible for
scattering Sgr~A* and the magnetar cannot be responsible for all of 
intense scattering towards the GC on this scale. 

The pulsar J1746$-$2856 at a separation $\simeq 0.2$~deg has a
  scattering location $\simeq 1-2$ kpc from the GC. In the two-component model this would be
caused by contributions from the local GC and distant
screens. Instead, it could be due to a separate, single
thin screen at $D_s \simeq 7$ kpc (figure~\ref{locmap}),
compatible with a location in the inner spiral arms as well as near the GC
region \citep[e.g., possibly the 3 kpc arm,][]{sannaetal2014}. Our observations cannot
distinguish between these possibilities. The weak constraint for PSR
J1746$-$2849 leaves it compatible with the scattering location of
either J1745$-$2912 or J1746$-$2856.

\subsection{Associations with known HII regions}
\label{sec:assoc-with-known}

We checked the lines of sight towards our pulsars against catalogs of HII
regions \citep{andersonetal2014,lockman1989} and supernova remnants
\citep[SNRs,][]{green2014} in the inner Galactic plane. The sources are shown compared to the 3D scattering
locations for each non-GC pulsar in Figure~\ref{candidatemap}. From the WISE catalog, we identify one promising
candidate HII region for each of B1809$-$176 and B1822$-$14 with
separation comparable to the measured radius \citep[S30 and S40,][]{sharpless1959}. Additionally, the line
of sight towards B1758$-$23 is close to ($\lesssim 5$ pc from) the
Trifid Nebula at a distance $\simeq 1.8$ kpc, which hosts an O star
with a large HII region \citep[S50,][]{lyndsetal1985,cordeslazio2003}. This
region may also be interacting with the SNR W28 at a distance $\simeq
2$ kpc \citep[e.g.,][]{velazquezetal2002}.

In all three cases, the candidate sources overlapping in the sky plane
are located at distances commensurate with the scattering locations
$D_s$ inferred for the pulsars. The association of the scattering
medium with these HII regions (and/or the SNR in the case of B1758$-$23)
therefore seems likely. For the GC sources, there are no known candidate HII regions on the
line of sight towards J1745$-$2912, while J1746$-$2849,  J1746$-$2856, and
SGR J1745$-$2900 are covered by at least one candidate HII region. However, claiming
associations between the HII regions and GC scattering is
difficult. Kinematic distances cannot be determined towards the GC,
and the region is crowded with sources both along the lines of sight
and within the GC region. 

Interstellar scattering has also been proposed to originate at the
ionized outer regions of giant molecular clouds. The line of sight to
two of the non-GC pulsars pass near the edge of the candidate HII
regions, possibly consistent with this scenario.

\subsection{The distance to B1758$-$23}
\label{sec:distance-b1758-23}

The extragalactic background source J1801-231 is only 2 arcminutes
from B1758$-$23 in angular separation. \citet{claussenetal2002} showed
that its angular size $\theta \sim \nu^{-2}$ or steeper, as
predicted for scatter-broadening. Assuming that B1758$-$23 and
J1801-231 share a single, thin scattering medium, those 
constraints along with the temporal broadening of the pulsar provide
unique solutions for the distance to the scattering medium and to the
pulsar:

\begin{eqnarray}
D_s &=& 2.292 \frac{\tau_p}{\theta_p \theta_{\rm ex}},\\
D &=& \frac{D_s}{1-\theta_p/\theta_{\rm ex}},
\end{eqnarray}

\noindent where $D$ and $D_s$ are in kpc, $\tau_p$ is in seconds, 
$\theta_{\rm ex}$ and $\theta_p$ are in arcseconds, and all quantities
are measured at a common frequency.

Using our measurement of the size of B1758$-$23, we repeat this
exercise. The quantities $\theta_{\rm ex}$ and $\tau_p$ have both been
measured up to a frequency $\nu \simeq 1.7$ GHz. We then need to
extrapolate our measured size back in frequency. In order to remain
consistent with our assumptions for measuring $D_s$ in \S
\ref{sec:locat-scatt-scre}, we scale $\theta_p^2 \propto \tau_p$,
using $\tau_p \propto \nu^{-\alpha}$ with $\alpha = 3.5 \pm 0.2$
(Table \ref{tautable}). At $1.7$ GHz, the values assumed are then
$\tau_p = 47 \pm 7$ ms, $\theta_{\rm ex} = 275 \pm 25$ mas, and $\theta_p
= 160 \pm 60$ mas. We again draw random Gaussian samples to measure a
new $D_s = 2.3 \pm 0.6$ kpc, consistent with the measurement
above. The extragalactic source size measured by
\cite{claussenetal2002} is therefore consistent with a shared
scattering origin with PSR B1758$-$23.

The difference between pulsar and extragalactic source sizes can then be used to estimate the distance to
the pulsar. The resulting value is $D = 5.3^{+1.4}_{-1.1}$~kpc,
consistent with the HI distance used in \S\ref{sec:locat-scatt-scre}. The probability distribution over $D$ is shown in Figure
\ref{b1758dist}. Large distances $\simeq 10$ kpc, as found in NE2001
due to the large pulsar DM, are disfavored at $> 2 \sigma$. The
measured distance rules out ($> 3 \sigma$) an association of PSR 
B1758$-$23 with the supernova remnant W28 at a distance $D \simeq 2$
kpc \citep[e.g.,][]{goss1968,arikawaetal1999,velazquezetal2002}. The
supernova remnant is at a distance compatible with $D_s$ and so could contribute to the observed scattering
(\S\ref{sec:assoc-with-known}).

\begin{figure*}
\begin{center}
\begin{tabular}{ll}
\includegraphics[scale=0.6]{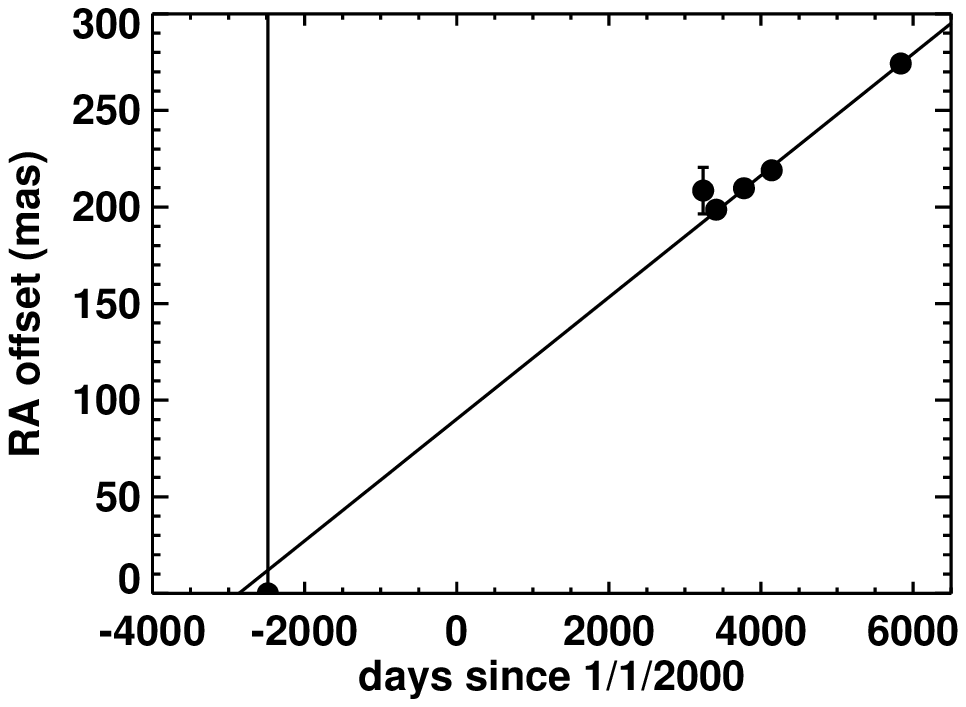}&
\includegraphics[scale=0.6]{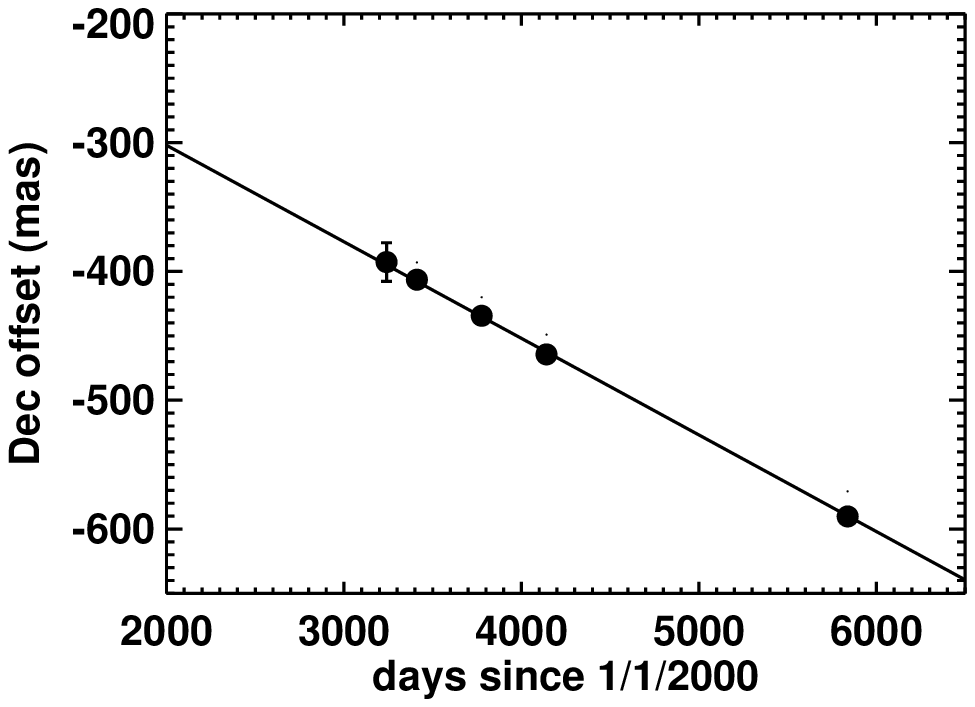}\\
\includegraphics[scale=0.6]{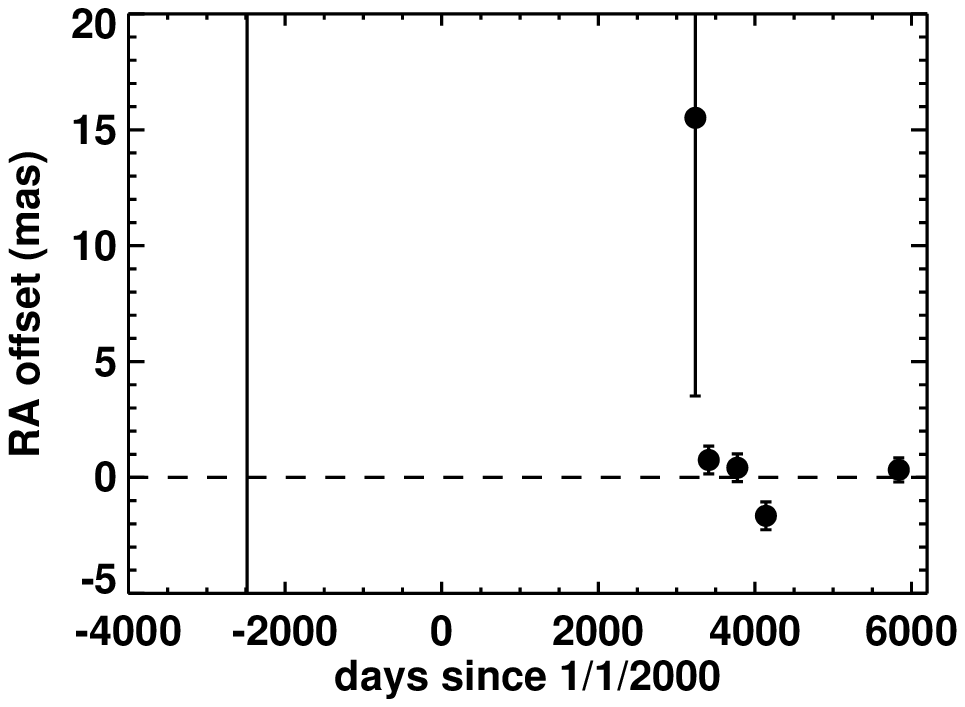}&
\includegraphics[scale=0.6]{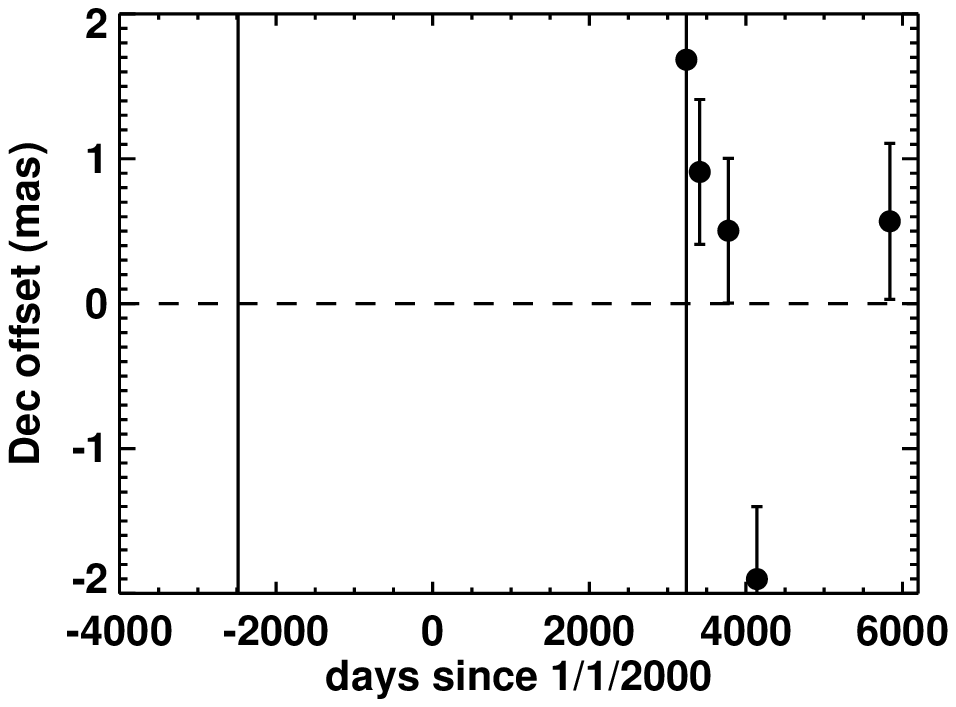}\\
\end{tabular}
\caption{\label{b1822propermotion}Proper motion of PSR B1822$-$14 in RA
  and Dec (top two panels) and the residuals of the VLBI data compared
to the best linear fit (bottom two panels). The first two epochs are
from archival VLA data \citep{frailscharringhausen1997}, the next four
are VLA and VLBA observations \citep{moldonetal2012}, and the final point is from our
recent VLBA observation. Our position is consistent with the proper
motion measured by \citet{moldonetal2012}.}
\end{center}
\end{figure*}

\subsection{B1822$-$14 proper motion}

\citet{moldonetal2012} used VLBA observations of B1822$-$14 at 5 GHz from
2009-2011 to
measure its proper motion to be $\mu_\alpha \cos{\delta} = 10.0 \pm 0.3
\hspace{2pt} \rm mas \hspace{2pt} yr^{-1}$, $\mu_\delta = -29.0 \pm
0.3 \hspace{2pt} \rm mas \hspace{2pt} yr^{-1}$. Figure
\ref{b1822propermotion} shows their data, archival VLA data 
\citep{frailscharringhausen1997}, and our new data point, along with fits for the
proper motion and the residuals. All VLBA observations used the common
phase reference source J1825$-$1718, and we correct all measurements to
reflect the latest measured calibrator position. To account for
systematics due to the different observing frequencies used (5 vs. 7.5
GHz), for example due to core shift of the calibrator source, an extra
$0.5$ mas error has been added to our data point. The
updated fit is

\begin{eqnarray}\nonumber
\alpha_{\rm J2000} = 18^{\rm h} 25^{\rm m} 2^{\rm s}.955067 \pm 0.29
 \pm 0.12 \hspace{2pt} \rm mas,\nonumber\\
\delta_{\rm J2000} =  -14^{\circ} 46' 53''.24531 \pm 0.26 \pm 0.17 \hspace{2pt} \rm mas,\nonumber\\
\mu_\alpha \cos{\delta} = 11.07 \pm 0.10 \hspace{2pt} \rm mas \hspace{2pt} \rm yr^{-1},\nonumber\\
\mu_\delta = -27.61 \pm 0.10 \hspace{2pt} \rm mas \hspace{2pt} \rm yr^{-1},\nonumber
\end{eqnarray}

\noindent where the weighted reference epoch is Oct. 20, 2011 (MJD
55854.7) and the second error terms are the current calibrator
position uncertainties in the ICRF. The fit result is poor: reduced
$\chi^2 = 4.1$. As can be seen from the residuals in figure
\ref{b1822propermotion}, there is a systematic $\simeq 2$ mas offset
between their final measurement and ours. This explains the discrepancy between the proper motion measurements at the
$\simeq 3-4 \sigma$ level. The residuals are likely 
due to underestimated systematics in comparing the positions, which
could result for example from a larger than average core shift or 
refractive image wander from scattering. On the
other hand, we can rule out their measured proper motion, since it
would lead to an offset of $\simeq 8$ mas in both RA and Dec from the
current position of PSR B1822$-$14. Future VLBI observations at 7.5
GHz, as used here, would allow a more robust proper motion measurement. In any case, we have verified the large proper motion
of this pulsar seen by \cite{moldonetal2012}. At the estimated DM distance of $\simeq 5.5$ kpc, this
corresponds to a space velocity of $\simeq 750 \, \mathrm{km} \, \mathrm{s}^{-1}$, confirmation
that B1822$-$14 is a runaway pulsar. 

\subsection{Large DM contributions from single HII regions and effect
  on distance estimates}

Assuming that the nearby HII regions above are responsible
for the observed temporal and angular broadening, we estimate the
minimum electron number density required to produce the observed
images at the inferred screen locations.

For scattering by a thin screen of material with a Kolmogorov
turbulent spectrum and inner/outer scales $L_{0,1}$, the image size is given
by \citep{vanlangeveldeetal1992}:

\begin{equation}
\theta = \frac{\pi \rho_C}{\sqrt{2 \ln 2} \lambda},
\end{equation}

\noindent where $\theta$ is the FWHM size as measured here,

\begin{equation}
\rho_C = \left[6\pi^2\lambda^2r_e^2\mathcal{L}(D)
  q_1^{1/3}\right]^{-1/2},
\end{equation}

\noindent and

\begin{equation}
\mathcal{L}(D) = \int_0^D dx \hspace{2pt} C_n^2 (x) \left(\frac{x}{D}\right)^2,
\end{equation}

\noindent where $C_n^2$ is the normalization of the turbulent power spectrum.

A lower limit to the average electron number density required to produce an observed
image size comes from assuming the density fluctuations are order
unity, $\delta n_e = n_e$, so that,

\begin{equation}
n_e \ge \delta n_e = (6 \pi C_n^2)^{1/2} \left(\frac{L_0}{2\pi}\right)^{1/3}.
\end{equation}

For the individual HII regions associated with these scattering
screens, we follow \citet{sichenederdexter2017} and set the line of sight distance through the cloud to
its measured radius, $R$, and the outer scale to $L_0 = f_2 R$, where
$f_2 \le 1$ is an unknown constant.

The density can then be written as,

\begin{equation}
n_e \simeq 130 \left(\frac{\theta}{10 \hspace{2pt} \rm mas}\right)
\left(\frac{\nu}{7.5 \hspace{2pt} \rm GHz}\right)^{-2}
\left(\frac{R}{3 \hspace{2pt} \rm pc}\right)^{1/6}\gamma^{-1}
f_2^{-1/3} \rm cm^{-3},
\end{equation}

\noindent where $\gamma = \Delta / D$ is the screen location and we
have assumed an inner scale of $L_1 = 10^4$ km \citep[e.g.,][]{wilkinsonetal1994}.

For the three non-GC pulsars with associated HII regions S30/S40/S50, we can
estimate the minimum contribution from the HII regions to the DM,
$\Delta$DM $= \eta \hspace{1pt} n_e R$, where $\eta = 0-2$ is the fraction of $R$
intersected by the line of sight. For parameter ranges of $f_2 = 1/10-1/2$,
$\eta = 1/2-1$, we find $\Delta \rm DM \simeq 250-1000$, $120-500$, $60-250$ \pcm for
B1758$-$23, B1809$-$176, and B1822$-$14 respectively. These lower limits are
$\simeq 25-100\%$ of the total DM in each case.

The Trifid nebula, lying close to the line of sight to B1758$-$23 and at
the distance we infer for the scattering, is ionized by the O7V star HD 164492A. For our $n_e
\simeq 130-220 \hspace{2pt} \rm cm^{-3}$
values for B1758$-$23 and the measured $R \simeq 4.5$ pc, we can use the
Str\"{o}mgren radius to 
calculate the photon flux, $N_{\rm Ly}$, required to ionize the HII region:

\begin{equation}
N_{\rm Ly} = 4/3 \pi R^3 n_e^2 \alpha_H \simeq 0.7-2.0 \times 10^{50} \hspace{2pt} s^{-1},
\end{equation}

\noindent consistent with expectations for this stellar type
\citep{sternbergetal2003}, and estimates based on the observed continuum radio
emission. The $n_e$ values we find also agree with
the measurement $n_e = 250 \pm 100$ \cm from line ratios
\citep{lyndsetal1985}.

Dispersion measures are frequently used to infer pulsar distances
\citep{taylorcordes1993,cordeslazio2002}. The NE2001 model includes contributions from
``clumps'' of electrons along many lines of sight, including
B1758$-$23. However, the contribution to the DM is assumed to be $\simeq
18$ \pcm \citep{cordeslazio2003}, a factor $\gtrsim 10$ smaller than
we infer would be the minimum contribution of the Trifid Nebula. These large DM
contributions could therefore significantly reduce the inferred distances to
pulsars located behind HII regions, an effect seen previously in
the Gum nebula \citep{johnstonetal1996}. Using the NE2001 model, we
calculate revised distances by subtracting a fiducial HII region
contribution of $50\%$ of the total DM. The new distances are $6.1$,
$3.8$, and $3.8$ kpc compared to NE2001 values of $12.6$, $6.2$, and
$5.1$ kpc. The revised distance estimate for B1758$-$23 of $6.1$ kpc is
comparable to that of $4 \pm 1$ kpc from HI absorption, and agrees 
with our estimate in \S
\ref{sec:distance-b1758-23}. The revised estimates are also in good
agreement with the new electron density model of \citet{yaoetal2017}. Including these large DM contributions from known HII
regions with density and radius could improve distance estimates for
some lines of sight through the Galactic plane.

\subsection{Systematic uncertainties}\label{sec:syst-uncert}

The measured sizes, extrapolated $\tau$ values, and source
distances are all subject to systematic errors, which could exceed the
statistical errors adopted in our analysis. Here we
briefly discuss how those errors could affect the results. 

\subsubsection{Phase calibration errors and source sizes}

The VLBA observations used phase referencing to nearby calibrator
sources. However, the intense scattering to
the Galactic plane often led to large offsets $\simeq 2-3^\circ$ to
the nearest suitable calibrator source, which was sometimes still
significantly scatter-broadened. Both effects can lead to 
residual phase errors which would cause us to overestimate the angular
broadening of the target sources. During the observations of PSR
J1745$-$2912, we also observed Sgr~A* using the same phase reference
calibrator.  The images of Sgr~A* formed without self-calibration were
broadened by $\sim$50\% compared to the expected size, which was
recovered after one round of self-calibration.  For the bright pulsar
B1822-14, self-calibration reduced the fitted size by $\sim$20\%.  We
could therefore expect the scatter-broadening of our other target
pulsars to be over-estimated by a similar factor, although the
different observing conditions on each day could lead to variations.

\subsubsection{Extrapolation of the measured pulse broadening}

Locating scattering screens requires measurements of the angular and
pulse broadening at the same frequency. In practice, the pulse
broadening can only be measured at lower frequencies than the VLBA+VLA
observations. As discussed in \S\ref{sec:temp-broad-dist}, we have
extrapolated archival pulse broadening data to the observed
frequencies. For sources with multiple measurements (J1746$-$2849, J1746$-$2856,
B1758$-$23, B1822$-$14), the errors from extrapolation are likely less
severe and can be estimated from the data. For J1745$-$2912 and B1809$-$176,
only one measurement is available and so these errors could be
larger. If we force $\alpha=4$ as predicted for the thin screen model,
all scattering locations move closer to Earth by $\approx
0.5-1$ kpc. In all cases except for B1822$-$14 and J1746$-$2856, this is within our $1
\sigma$ uncertainty region. For the GC magnetar J1745$-$2900, $\alpha = 3.8
\pm 0.2$, while for B1822$-$14 $\alpha = 3.8 \pm 0.3$. Those values
are consistent with either $\alpha = 4$ or $3.5$. Measuring the
temporal broadening more accurately would allow for a more
robust inference of the scattering location.

\subsubsection{Distance estimates}

We have assumed that the GC pulsars are located at the GC distance of $8.3$
kpc, as suggested by their high DM and $\tau$ values. The DM distance
for B1758$-$23 is much larger than estimates from HI absorption and scattering (figure
\ref{b1758dist}). For the other two sources, only DM distances are available. Those
distances could therefore be subject to systematic error of up to $\approx 50\%$.

\subsubsection{Possible effect on results}

The phase calibration errors likely cause us to overestimate the
angular broadening, $\theta$. Extrapolation of the pulse broadening
could cause us to underestimate $\tau$ for PSR J1745$-$2912 and
B1809$-$176, if $\alpha < 4$. For the other sources, if the thin
screen model holds and $\alpha = 4$, the $\tau$ values at high
frequency could be overestimated. DM distances could be systematically off
in either direction, but might be more likely to be overestimated
(e.g. as for the GC pulsars and maybe B1758$-$23).

There are two relevant limits from equation \eqref{eq:1}, $\Delta / D
\ll 1$ and $\Delta / D \approx 1$, for scattering local to and far
from the source. In the former limit, $\Delta \sim D^2 \theta^2 /
\tau$. In this case, overestimating $D$ or $\theta$ has the same
effect as underestimating $\tau$: all of these errors would cause us
to overestimate $\Delta$. In the opposite limit, the dependence is
most simply written as $D_s \sim \tau / \theta^2$. The distance to the
scattering screen is roughly independent of source
distance. Underestimating $\tau$ or overestimating $\theta$ would
still mean that the screen is located closer to the source than we infer. 

We simulate these effects using B1758$-$23 as an example, where $D_s =
2.0 \pm 0.5$~kpc using our best measurements. We choose this source
because the adopted errors are relatively small, but it is low
declination with evidence of phase calibration errors and has 
discrepant distance and $\tau$ measurements. If we were to
use $\alpha = 4$, the extrapolated $\tau$ value could be $\approx
60\%$ smaller, larger than our adopted error bar. Then we would find $D_s =
1.5 \pm 0.4$ kpc. Repeating the exercise for a $20\%$ smaller size or the larger DM distance, we find
$D_s = 2.3 \pm 0.6$ kpc and $D_s = 2.7 \pm 0.8$ kpc. Each individual
effect systemtically shifts $D_s$ at roughly the $1\sigma$
level.

Given the various possible systematic uncertainties, it is worth
considering how the analysis presented here could be improved in the
future. The low significance of our source detections leads to relatively poor
size measurements. Higher signal-to-noise
could allow phase self-calibration, reducing systematic
calibrator errors as well as the statistical error bars on the
sizes. Constraining image anisotropy would also allow us to test whether the same
scattering medium could be responsible for the images of both
J1746$-$2849 and J1746$-$2856, for example. This could be achieved
with longer integrations (all sources), by going to lower frequency
where the pulsars are brighter (compact sources, although the phase
calibrators may be heavily scatter-broadened), or by including the VLA
in observations of non-GC pulsars. A better measurement of the
frequency-dependence of the temporal broadening ($\alpha$) might be
even more important given the systematic uncertainties from
extrapolation. This is also difficult for faint sources, especially at
higher frequencies where the temporal broadening becomes much smaller
than the intrinsic pulse width. We have also assumed a single, thin
screen model in order to measure $D_s$. At low frequencies
the shallow temporal broadening slopes $\alpha < 4$ imply that this
assumption breaks down. At the higher frequencies of our VLBA+VLA
observations the thin screen approximation may still be valid
\citep{cordeslazio2001}. This assumption could be
tested by measuring the frequency-dependence of both $\theta$ and
$\tau$, or for very bright sources through phase-resolved
imaging. Both techniques have been used to show that the Sgr A* and GC
magnetar images are likely dominated by a single thin screen
\citep{boweretal2014,spitleretal2014,wucknitz2015}. 

\section{Summary}
\label{sec:summary}

We have used VLBA+VLA observations to measure the scatter-broadened
image sizes of 6 of the most heavily scattered known
pulsars. Combining the image sizes with previously measured
temporal broadening of the pulse profiles leads to an estimate of the
location of the scattering medium along the line of sight. A summary
of our results follows.

\begin{itemize}
\item Three of the nearest pulsars to the GC
  magnetar SGR J1745$-$2900 have smaller image sizes despite comparable
  temporal scattering, evidence of additional strong scattering component closer to or even within
  the GC region. The strength of the Sgr~A* / magnetar scattering
  screen decreases by at least a factor of 5 on scales of $\simeq 10$
  pc. The GC region shows significant scatter-broadening on larger scales of
  $\simeq 100$ pc, which has contributions from at least two distinct
  sources \citep[the screen responsible for the scattering of Sgr A* and SGR
  J1745$-$2900,][and an additional screen $\lesssim 2$~kpc from the GC]{boweretal2014}. The temporal broadening for all known GC pulsars is comparable
  $\simeq 1$ s at 1 GHz, $\lesssim 10^{2-3}$ orders of magnitude
  weaker than the proposed ``hyperstrong'' scattering medium in the
  GC. Nonetheless, the
  variable strength and locations of the scattering could imply
  variability in the temporal broadening with location and/or time,
  potentially reducing the
  sensitivity of past surveys to detecting short period (especially
  millisecond) pulsars.
\item We tentatively measure a very compact size $\simeq 2$ mas for
  the GC pulsar J1745$-$2912. Combined with its large degree of temporal
  broadening, this measurement locates the scattering to $\lesssim
  700$ pc of the source, likely within the GC region itself. 
\item The three non-GC pulsars in our sample all show scattering media
  located $\simeq 2$ kpc from Earth, likely within the
  Carina-Sagittarius spiral arm. In addition, all three have 3D
  positions consistent with known HII regions (S30, S40, S50, and in the case of
  B1758$-$23 also the supernova remnant W28). 
\item Assuming the likely association of the observed scattering with
  these HII regions, we calculate the minimum electron density
  required for them to produce the observed scattering. The
  corresponding minimum DM contribution is a large fraction $\gtrsim
  25\%$ of the total, suggesting that distances to these pulsars, and
  others lying behind HII regions, based on their DM could be significantly over-estimated.
\item Following \citet{claussenetal2002}, we independently constrain 
  the distance to B1758$-$23 as $D = 5.3^{+1.4}_{-1.1}$~kpc based on its temporal and angular
  broadening and the angular broadening of a very nearby extragalactic background source. This
  distance agrees with both our revised distance estimate of
  $\simeq 6.1$ kpc from including the HII region DM contribution and a
  measurement from HI absorption of $D = 4 \pm 1$ kpc. It rules out an
  association of the pulsar with the supernova remnant W28.
\item Finally, we have measured a new position of B1822$-$14, a known
  runaway pulsar \citep{moldonetal2012}. Our added epoch further
  constrains the proper motion and confirms previous measurements of a
  space velocity $\simeq 750 \, \mathrm{km} \, \mathrm{s}^{-1}$.

\end{itemize}

\section*{acknowledgements}
We thank the WISE team, \citet{cordeslazio2002}, and \citet{green2014} 
for making their data publicly available and D.H.F.M. Schnitzeler for
sharing ATCA position and flux density measurements for the GC pulsars
studied here. JD was supported by a Sofja
Kovalevskaja Award from the Alexander von Humboldt Foundation of
Germany, and in part by the National Science 
Foundation under Grant No. NSF PHY-1125915. M. Kramer acknowledges 
financial support by the European Research Council for the ERC Synergy
Grant BlackHoleCam under contract no. 610058. L.G.S. gratefully
acknowledges support from the ERC Starting Grant BEACON under contract
no. 279702 and the Max Planck Society. The Parkes radio telescope is
part of the Australia Telescope, which is funded by the Commonwealth
Government for operation as a National Facility managed by CSIRO. The
National Radio Astronomy Observatory is a facility of the National
Science Foundation operated under cooperative agreement by Associated
Universities, Inc.

\bibliographystyle{mnras}
\bibliography{pulsars}

\bsp	\label{lastpage}
\end{document}